\documentclass[conference]{IEEEtran}
\usepackage{cite}
\usepackage{graphicx}

\usepackage{anyfontsize}

\usepackage[many]{tcolorbox}

\usepackage{mathtools,latexsym,stmaryrd}
\usepackage{pbox}
\usepackage{xfrac}
\usepackage{booktabs}
\usepackage{colortbl}
\usepackage{threeparttable}
\usepackage{amsthm}
\usepackage{scalerel}
\usepackage{multirow}
\usepackage{tikz}
\usepackage{mathrsfs}
\usepackage{mathpartir}
\usepackage{algorithm}
\usepackage{url}
\usepackage{syntax}
\usepackage{framed}
\usepackage[noend]{algpseudocode}
\usepackage{flushend}
\usepackage{placeins}
\usepackage{listings}
\usepackage{moresize}
\usepackage{wrapfig}
 
\usepackage{seqsplit}
\usepackage{alltt}
\usepackage{xspace}
\usepackage{soul}
\usepackage{bm}
\let\labelindent\relax
\usepackage{enumitem}
\usepackage{pifont}

\newcommand{\reffig}[2]{\F~\hyperref[#1]{#2}}

\DeclareMathAlphabet{\mathcal}{OMS}{cmsy}{m}{n}

\usepackage{amsmath, listings, amsthm, amssymb, proof, xspace}
\usepackage[inference]{semantic}

\RenewDocumentCommand\Pr{sO{}r()}{%
  \operatorname{Pr}%
  \begingroup
  \IfBooleanTF{#1}
    {\PrInn*{#3}}
    {\PrInn[#2]{#3}}%
  \endgroup
}

\DeclarePairedDelimiterX\PrInn[1](){%
  \activatebar
  #1%
}

\newcommand{\activatebar}{%
  \begingroup\lccode`~=`|
  \lowercase{\endgroup\def~}{\;\delimsize\vert\;}%
  \mathcode`|=\string"8000
}

\usepackage{array}
\newcommand{\PreserveBackslash}[1]{\let\temp=\\#1\let\\=\temp}
\newcolumntype{C}[1]{>{\PreserveBackslash\centering}p{#1}}
\newcolumntype{R}[1]{>{\PreserveBackslash\raggedleft}p{#1}}
\newcolumntype{L}[1]{>{\PreserveBackslash\raggedright}p{#1}}

\usepackage{thmtools}

\declaretheoremstyle[spaceabove=\topsep,notefont=\normalfont\itshape]{mystyle}

\usepackage{footnote}
\makesavenoteenv{tabular}
\makesavenoteenv{table}

\usetikzlibrary{arrows,patterns, decorations.pathreplacing}

\newcommand{\F}{Fig.}

\definecolor{ForestGreen}{RGB}{34,139,34}
\definecolor{matplotlib-purple}{HTML}{9467bd}
\definecolor{matplotlib-blue}{HTML}{1f77b4}
\definecolor{matplotlib-green}{HTML}{2ca02c}
\definecolor{matplotlib-orange}{HTML}{ff7f0e}
\definecolor{matplotlib-red}{HTML}{d62728}
\definecolor{matplotlib-brown}{HTML}{8c564b}
\definecolor{matplotlib-pink}{HTML}{e377c2}
\definecolor{matplotlib-gray}{HTML}{7f7f7f}
\definecolor{matplotlib-yellow}{HTML}{bcbd22}
\definecolor{matplotlib-cyan}{HTML}{17becf}

\definecolor{my-orange}{RGB}{232, 98, 88}

\newcommand{\parh}[1]{\noindent\textbf{#1}}

\newcommand{\ignore}[1]{}

\lstdefinestyle{base}{
  moredelim=**[is][\color{red}]{@}{@},
  escapeinside={<@}{@>}
}

\usepackage{amssymb}
\usepackage{pifont}

\newcommand\DejaVuttfamily{%
  \fontfamily{DejaVuSansMono-TLF}\selectfont }

\lstdefinestyle{base}{
  moredelim=**[is][\color{red}]{@}{@},
  escapeinside={<@}{@>}
}

\lstdefinelanguage
   [x64]{Assembler}     
   [x86masm]{Assembler} 
   {morekeywords={CDQE,CQO,CMPSQ,CMPXCHG16B,JRCXZ,LODSQ,MOVSXD, %
                  POPFQ,PUSHFQ,SCASQ,STOSQ,IRETQ,RDTSCP,SWAPGS, %
                  rax,rdx,rcx,rbx,rsi,rdi,rsp,rbp, %
                  r8,r8d,r8w,r8b,r9,r9d,r9w,r9b}} 

\usepackage{listings}
\usepackage{fancyvrb}
\usepackage{color}
\definecolor{lightgray}{rgb}{.9,.9,.9}
\definecolor{darkgray}{rgb}{.4,.4,.4}
\definecolor{purple}{rgb}{0.65, 0.12, 0.82}
\definecolor{commentgreen}{RGB}{63,127,95}

\colorlet{myPurple}{blue!40!red}
\definecolor{myOrange}{RGB}{255,192,0}

\newcommand{\enc}[1]{$\phi^{*}_{\theta}$}
\newcommand{\dec}[1]{$\psi^{*}_{\theta}$}

\lstdefinelanguage{Solidity}{
  keywords={len,delete,int,void,payable, public, event, contract, typeof, new, true, false, catch, function, return, null, catch, switch, var, if, in, while, do, else, case, break,struct,const,socklen_t,sa_familty_t,char,sockaddr},
  keywordstyle=\color{violet}\bfseries,
  ndkeywords={class, export, boolean, throw, implements, import, this},
  ndkeywordstyle=\color{darkgray}\bfseries,
  identifierstyle=\color{black},
  sensitive=false,
  comment=[l]{//},
  escapeinside={(*@}{@*)},          
  morecomment=[s]{/*}{*/},
  commentstyle=\color{commentgreen}\ttfamily,
  stringstyle=\color{red}\ttfamily,
  morestring=[b]',
  morestring=[b]"
}

\newcommand{\rnum}[1]{\uppercase\expandafter{\romannumeral #1\relax}}

\algnewcommand{\LeftComment}[1]{\Statex \(\triangleright\) #1}

\definecolor{pptbrown}{RGB}{132,60,12}
\definecolor{pptgreen}{RGB}{56,87,35}

\let\OLDthebibliography\thebibliography
\renewcommand\thebibliography[1]{
  \OLDthebibliography{#1}
  \setlength{\parskip}{0pt}
  \setlength{\itemsep}{0pt plus 0.1ex}
}

\definecolor{pptgreen}{RGB}{84,130,53}
\definecolor{pptred}{RGB}{176,35,24}

\definecolor{pptgreen1}{RGB}{78,173,91}
\definecolor{pptred1}{RGB}{192,0,0}

\definecolor{pptyellow1}{RGB}{203,195,167}
\definecolor{pptgreen2}{RGB}{184,192,176}

\definecolor{pptred3}{RGB}{192,0,0}
\definecolor{pptyellow3}{RGB}{255,192,0}
\definecolor{pptgreen3}{RGB}{4,216,178}

\definecolor{pptblue}{RGB}{0,176,240}
\definecolor{pptgrey}{RGB}{175,171,171}

\newlength{\dpcircle}
\newlength{\rcircle}
\newlength{\dcircle}

  \DeclareFontFamily{U}{dutchcal}{\skewchar \font =45}
  \DeclareFontShape{U}{dutchcal}{m}{n}{
    <-> dutchcal-r}{}
  \DeclareFontShape{U}{dutchcal}{b}{n}{
    <-> dutchcal-b}{}
  \DeclareMathAlphabet{\mdutchcal}{U}{dutchcal}{m}{n}
  \SetMathAlphabet{\mdutchcal}{bold}{U}{dutchcal}{b}{n}
  \DeclareMathAlphabet{\mdutchbcal} {U}{dutchcal}{b}{n}

  \DeclareFontFamily{U}{txcal}{\skewchar \font =45}
  \DeclareFontShape{U}{txcal}{m}{n}{
    <-> txr-cal}{}
  \DeclareFontShape{U}{txcal}{b}{n}{
    <-> txb-cal}{}
  \DeclareMathAlphabet{\mtxcal}{U}{txcal}{m}{n}
  \SetMathAlphabet{\mtxcal}{bold}{U}{txcal}{b}{n}
  \DeclareMathAlphabet{\mtxbcal} {U}{txcal}{b}{n}



\usepackage[hidelinks]{hyperref}

\usepackage[capitalise]{cleveref}

\definecolor[named]{ACMBlue}{cmyk}{1,0.1,0,0.1}
\definecolor[named]{ACMYellow}{cmyk}{0,0.16,1,0}
\definecolor[named]{ACMOrange}{cmyk}{0,0.42,1,0.01}
\definecolor[named]{ACMRed}{cmyk}{0,0.90,0.86,0}
\definecolor[named]{ACMLightBlue}{cmyk}{0.49,0.01,0,0}
\definecolor[named]{ACMGreen}{cmyk}{0.20,0,1,0.19}
\definecolor[named]{ACMPurple}{cmyk}{0.55,1,0,0.15}
\definecolor[named]{ACMDarkBlue}{cmyk}{1,0.58,0,0.21}

\usepackage{microtype}

\usepackage[normalem]{ulem}  

\begin{document}
%

\title{ZKCache: Cache Side-Channel Attacks on
Zero-Knowledge Machine Learning Systems}

\author{\IEEEauthorblockN{Sen Deng\IEEEauthorrefmark{1},
		Zhibo Liu\IEEEauthorrefmark{2},
		Yuanyuan Yuan\IEEEauthorrefmark{3},
		Zheng Zhang\IEEEauthorrefmark{4}, and
		Shuai Wang\IEEEauthorrefmark{1}}
	\IEEEauthorblockA{\IEEEauthorrefmark{1}The Hong Kong University of Science and Technology\\
		Emails: sdengan@cse.ust.hk, shuaiw@cse.ust.hk}
	\IEEEauthorblockA{\IEEEauthorrefmark{2}State Key Laboratory of Novel Software Technology, Nanjing University\\
		Email: zhiboliu@nju.edu.cn}
	\IEEEauthorblockA{\IEEEauthorrefmark{3}Tsinghua University\\
		Email: yyyuan@mail.tsinghua.edu.cn}
	\IEEEauthorblockA{\IEEEauthorrefmark{4}Southern University of Science and Technology\\
		Email: aca2hang2heng@gmail.com}}
\maketitle

\begin{abstract}
Zero-knowledge machine learning (ZKML) lets a prover convince a verifier that a
neural network was evaluated correctly while keeping the model, the input, or
both private.  A growing use case is protecting a \emph{proprietary} model inside
a hosted proving service, where the architecture itself is the confidential
asset.  Zero knowledge, however, protects only the proof transcript. Proof
generation is ordinary software, and its memory-access behavior during proving
depends on the model it is meant to hide.  An unprivileged process that shares a CPU with the prover, with no
shared memory and no privileged performance counters, can recover the
proprietary architecture purely from cache contention during proof generation.

Our attack, ZKCache, pairs cache-contention measurements with a source-level
model of how a network's shape drives the prover's work. It combines learned
trace evidence with exact structural constraints to recover operator types, their
hyperparameters, and the network's computation graph.  We evaluate two
open-source Halo2-based ZKML systems, EZKL and TeleSparse, on VGG16 and
ResNet-18.  ZKCache recovers operator types with 98.6\% accuracy and
hyperparameters with 93.2\% accuracy. For ResNet-18, it recovers the exact
topology, meaning the complete operator-type sequence together with all
residual connections, in 83.3\% of trials.  We further show that the leak is
mitigable. A compiler defense that groups shape-compatible layers under a
developer-specified budget hides per-layer parameters at 53\%
end-to-end overhead.  ZKCache demonstrates that transcript-level zero knowledge
does not, by itself, keep a model's architecture~secret.
\end{abstract}
\IEEEpeerreviewmaketitle



	\vspace{-3pt}
\section{Introduction}
\label{sec:introduction}

A zero-knowledge proof lets one party convince another that a computation was
performed correctly while revealing nothing beyond the statement being
proved~\cite{10.1145/22145.22178}.  Zero-knowledge machine learning (ZKML)
applies this guarantee to neural-network inference. A prover demonstrates that a
specific model produced a claimed output on a given input without disclosing the
private parts of the computation, which may be the input, the model parameters,
or both.  Modern systems such as EZKL~\cite{ezkl} and
TeleSparse~\cite{telesparse2025} compile increasingly complex networks into
arithmetic circuits and are already used to prove inference for real
models~\cite{chen2024zkml,ezkl,telesparse2025}.

A prominent deployment keeps the \emph{model itself} private. A service holds a
proprietary, expensively trained network and offers proofs that it ran
correctly, while treating the architecture, its operators, and their
hyperparameters as confidential intellectual property.  The model, its weights,
the compiled circuit, the proving key, and the witness all remain on the server,
and a client learns only the proof and the public statement.  The security
property at issue is \emph{architecture confidentiality}, namely whether ZKML
keeps the shape of a proprietary model secret.

Zero knowledge is a property of the protocol \emph{transcript}. It bounds what
the exchanged messages reveal and says nothing about how the prover computes
them.  Yet proof generation is performed by a concrete program with model-shaped
work, and the amount and pattern of that work depend on the architecture being
proved.  On shared hardware this execution becomes an attack surface distinct
from the protocol itself, namely the prover's memory behavior while it produces
the proof, not the proof it ultimately sends.  A process co-resident with
the prover can measure that behavior through the CPU cache without ever seeing
the model, its weights, the circuit, the proving key, or the witness.  We ask
the following question.

\begin{quote}
\emph{
Can ZKML safeguard model architecture privacy against co-resident cache side-channel attacks during proof generation?}
\end{quote}

Cache and related side channels have already been used to steal DNN
architectures, but only from ordinary inference.  Prior work recovers layer
sequences and tensor dimensions by matching noisy traces against profiled
operators~\cite{hong2018security,deepcache2024,hong20200wn,cachetelepathy2020,
ganred2020,deepsniffer2020}, exploits electromagnetic or accelerator-bus
emanations~\cite{batina2019csinn,yu2020deepem,zhu2021hermes}, or reconstructs
models from compiled executables~\cite{wu2022dnd,liu2023decompiling}.  These
attacks depend on shared framework or library pages that expose function-level
events, on physical or privileged observation interfaces, or on behavior
specific to one inference stack or compiler~\cite{hong20200wn,
cachetelepathy2020,deepsniffer2020,deepcache2024}.  None of these conditions
holds for a ZKML prover that shares only the cache.

ZKML execution is also structurally harder to read.  Ordinary inference
evaluates each operator once as a stream of numerical kernels, whereas proving
executes the model in different forms.  The prover first computes the values that
become the witness and later recomputes the same tensors while constructing the
circuit.  Tensor access and circuit-cell assignment interleave at fine
granularity during this second model-dependent stage.  The cryptographic
backend then commits the completed circuit and constructs the proof.
Recovering a model from such a trace poses three challenges that
ordinary-inference attacks do not~face.

\begin{itemize}[leftmargin=*,topsep=2pt,itemsep=1pt]
  \item \textbf{Phase isolation.}  The different stages of model-dependent
  computation must be distinguished from one another and separated from the
  architecture-independent cryptographic work that follows them.
  \item \textbf{Shape-to-memory attribution.}  Private properties such as kernel
  size, channel count, and stride must be inferred from coarse, noisy cache-set
  contention rather than read directly.
  \item \textbf{Structure beyond a layer sequence.}  For residual networks,
  layer-order recovery alone cannot reconstruct the model architecture.
  Identity and projection shortcuts leave minimal standalone cache footprints
  yet define the underlying graph~topology.
\end{itemize}

Because these systems are open source, their code specifies the relationship
between a model's shape and the prover's memory work. In EZKL, one proof
invocation has two cache-visible, model-dependent phases. The \emph{dummy-layout
view} computes the model's tensor values, and the \emph{synthesis-composite
view} commits them into the circuit. A layer's hyperparameters determine the
memory work in both phases.
These relations locate the model-dependent signal and give the attack
checkable constraints for interpreting a noisy trace. We develop them in
Section~\ref{sec:leakage-analysis}.

Building on this analysis, ZKCache makes side-channel recovery
\emph{prover-aware}. It combines learned cache-trace evidence with analytical
constraints derived from the prover's public implementation and structural
constraints from the network graph. The prover code is analyzed offline at design
time to derive the constraints. The attack
itself observes only cache contention and needs no access to the prover's source,
the model, its weights, or its memory, nor any shared memory or privileged
performance counter.  We develop its encoder, retrieval, and constraint solver in
Section~\ref{sec:attack-design}.

We evaluate ZKCache end to end on VGG16~\cite{simonyan2014very} and
ResNet-18~\cite{he2016deep}, each run on EZKL and TeleSparse, giving four
system--model pairs.  From single traces, it recovers operator types with 98.6\%
accuracy and their hyperparameters with 93.2\% accuracy.  Reconstructing a
residual network's exact topology is the hard case, since identity and projection
shortcuts must also be inferred.  Our all-or-nothing topology metric credits a
trial only when the complete operator-type sequence and every graph edge match;
it does not require every hyperparameter tuple to be covered.  ZKCache recovers
ResNet-18's exact topology in 83.3\% of trials.

In summary, this paper makes the following contributions.

\begin{itemize}[leftmargin=*,topsep=0pt,itemsep=2pt]
    \item \textbf{A new attack surface.}  A ZKML
    prover's \emph{execution} leaks the proprietary model architecture even when
    the proof transcript is zero knowledge.  We identify two model-dependent
    temporal views in Halo2-based proof generation and show that, within the
    synthesis-composite view, tensor access and circuit-cell assignment interleave
    as sources of leakage.  Transcript zero knowledge is necessary but not sufficient for
    architecture-private ZKML.

    \item \textbf{A prover-aware recovery attack.}  We design ZKCache, which
    learns one trace representation per temporal view and resolves the resulting
    candidates jointly under the analytical and graph constraints.  The
    combined evidence recovers operator types,
    hyperparameters, and the full
    computation graph from noisy cache contention, without shared memory or
    privileged counters.

    \item \textbf{End-to-end evaluation.}  We demonstrate recovery on
    two independent Halo2-based systems, EZKL and TeleSparse, and two
    architectures, VGG16 and ResNet-18, reaching 98.6\% operator-type and 93.2\%
    hyperparameter accuracy and fully reconstructing ResNet-18's residual graph.

    \item \textbf{A practical defense.}  We design and evaluate budgeted
    block-level envelopes, implemented by a compiler pass that groups shape-compatible
    convolutions into fixed, oblivious profiles under developer-specified
    performance caps, so that a trace reveals only a coarse profile rather than
    exact per-layer parameters.  A TeleSparse VGG16 prototype incurs
    53\% end-to-end overhead.
\end{itemize}

	\section{Background}
\label{sec:background}

A zero-knowledge proof conceals the underlying execution values but does not
require an implementation to hide its computation structure. In ZKML, the prover
fills an arithmetic circuit by executing the target model's operators. The model
topology consequently determines the memory-access pattern during proof
generation, even when the final proof satisfies the zero-knowledge property. We
first describe the ZKML setting and the scope of its guarantee
(Section~\ref{sec:background-zkml}). We then explain how a Halo2 prover turns
model operators into memory work (Section~\ref{sec:background-halo2}) and how
EZKL runs the model twice within one proof (Section~\ref{sec:background-systems}).
We next describe the cache side channel at cache-set granularity
(Section~\ref{sec:background-cache}) and define the model information treated as
secret (Section~\ref{sec:background-model-information}).

\subsection{Zero-Knowledge Machine Learning}
\label{sec:background-zkml}

A zero-knowledge proof lets a prover convince a verifier that a public statement
$s$ admits a witness $w$ with $\mathcal{R}(s,w)=1$, while revealing nothing
about $w$ beyond its existence~\cite{10.1145/22145.22178}.  In ZKML the relation
$\mathcal{R}$ encodes a neural-network inference. The public statement carries
the claimed output and commitments to any hidden object, and the witness carries
whatever the deployment keeps private~\cite{chen2024zkml,ezkl,telesparse2025}.  We
study the model-private setting, in which that private object is the network,
its operator graph, and the operators' hyperparameters.
Section~\ref{sec:background-model-information} makes this target precise.

Neural networks are normally evaluated over floating-point tensors, whereas
the arithmetic circuit operates over a finite field.  A ZKML compiler
quantizes tensors, lowers graph operators to field arithmetic and
lookup constraints, and fixes a circuit shape large enough for the resulting
computation~\cite{feng2021zen}.  Proof generation then has two conceptually
distinct inputs, a \emph{circuit description}, which defines the permitted
computation, and a
\emph{witness}, which supplies values for one execution.  Setup derives the
proving and verification keys for the circuit; proving fills the circuit with
witness-dependent values and creates a proof; verification checks
that proof against the public statement.

The zero-knowledge property concerns what can be learned from the protocol
messages.  It does not, by itself, require the prover implementation to use
the same instruction sequence or memory addresses for every possible private
model.  Consequently, an implementation can produce a zero-knowledge proof
while retaining model-dependent timing and memory behavior.  ZKCache studies
this distinction between transcript privacy and implementation-level
obliviousness.

\subsection{Halo2 Proof Generation}
\label{sec:background-halo2}

We focus on Halo2~\cite{halo2book}, a zero-knowledge proving system. Its PLONKish
arithmetization~\cite{cryptoeprint:2019/953} encodes a computation as constraints
over a table of cells and underlies widely used ZKML toolchains, including EZKL
and TeleSparse. The execution structure below is shared by PLONKish provers that
rebuild the model during proving, allowing the analysis to extend beyond one
codebase (Section~\ref{sec:discussion}).

\parh{Circuits as tables.}
A PLONKish circuit is a table of $n=2^k$ rows and a fixed set of
columns~\cite{cryptoeprint:2019/953,halo2book}.  \emph{Advice} columns hold the
values the prover computes and fills in, and are the only part of the table that
depends on the private witness; \emph{fixed} columns hold circuit constants and
the selectors that switch gates on and off.  Gates constrain a handful of cells
in the same or neighboring rows, and \emph{lookups} require a tuple of cells to
appear in a reference table, which is how a prover cheaply encodes nonlinear
operators and range checks that plain arithmetic would express
poorly~\cite{cryptoeprint:2020/315,halo2book,hao2024nonlinear}.  A \emph{gadget} is a
reusable bundle of gates and lookups that encodes one operation, such as a dot
product or a comparison; each model operator is lowered to one or more gadgets.

\parh{Synthesis turns operators into cells.}
Producing a proof has two stages that touch different memory.  In
\emph{circuit synthesis}, a layouter walks the model's operators in order and,
for each one, reads and reshapes the tensor slices it needs and then writes the
resulting operands, accumulators, and selector state into advice cells.  Reads
and writes are thus interleaved in one timeline, not split into separate read
and write stages.  A model tensor is not stored as one block. Its
elements and intermediate values are scattered across many advice cells
according to the gadget each operator uses, so the number of cells written and
the order in which they are touched are fixed by the model's shape. Larger
kernels, more filters, or a different stride change how much the prover assigns
and in what pattern.

\parh{Cryptographic proving.}
Once the table is filled, \emph{cryptographic proving} treats each column as a
polynomial, blinds and commits the witness-dependent ones, and builds the
permutation, lookup, and quotient arguments that show every constraint
holds~\cite{halo2book}.  Both stages run inside one proving call, but they touch
different in-memory objects and so leave different cache footprints.  The
shape-dependent leakage we analyze arises in synthesis and in the sizing pass
described next, not in this polynomial stage, whose memory footprint tracks the
padded circuit size rather than fine model dimensions.

\subsection{Target ZKML Systems}
\label{sec:background-systems}

We study two Halo2-based systems, EZKL and TeleSparse.  EZKL is our source-audit
target. Its open implementation exposes the path from an ONNX model to Halo2 cell
assignments, allowing us to trace how each model dimension produces repeated
memory work.  TeleSparse builds on the same arithmetization with
different optimizations, and serves to test whether the leakage survives a
change of codebase.  We run both systems on the same two architectures, VGG16
and ResNet-18.

\parh{EZKL.}
EZKL compiles an ONNX graph into a Halo2 circuit and then runs the model in
memory \emph{twice} within a single end-to-end proof, which is the origin of the
two temporal views our attack separates.

The first execution sizes the circuit. Before it can prove anything, EZKL must
know how many rows and cells the model needs, so it runs every operator's layout
over the real input and weight tensors while only counting the cells that would
be written, without populating a Halo2 table.  This sizing pass still performs
the model-dependent tensor traversal, for example sliding a convolution window
across the feature map, and we call the resulting trace the
\emph{dummy-layout view}.

The second execution produces the proof. The witness EZKL saves from the first pass records
the inputs, outputs, and range metadata, but not the intermediate activations,
so proving must recompute them. It replays the same operator layouts, now
writing real advice cells.  Because tensor reads and cell assignments
are interleaved here (Section~\ref{sec:background-halo2}), we call this second
trace the \emph{synthesis-composite view}.  The two views therefore arise from the
same shape-dependent computation, once while sizing and once while assigning.
Section~\ref{sec:leakage-analysis} traces this behavior to the specific source
routines and derives what each view leaks about a model dimension.

\parh{TeleSparse.}
TeleSparse lowers proving cost with model sparsification and neural
teleportation~\cite{telesparse2025}.  These optimizations differ from EZKL's
ONNX lowering, but its prover still expresses inference as PLONKish rows,
columns, gates, and lookups, and still rebuilds the model during synthesis.  We
instantiate VGG16 and ResNet-18 in both systems and build a separate shadow
corpus for each, which lets us test the attack across codebases under controlled
conditions.

\subsection{Cache Side Channels}
\label{sec:background-cache}

Modern processors divide memory into fixed-size cache lines and map each line
to a set in a set-associative cache.  Processes that share a cache compete for
these sets, so one process can infer another process's cache use by measuring
its own access latency.  A private L1 cache generally requires the adversary
to share a physical core with the victim, whereas a shared last-level cache
can permit observations across cores on the same processor
package~\cite{liu2015last}.

Prime+Probe does not require the attacker and victim to share memory
pages~\cite{osvik2006cache,liu2015last,oren2015spy}.  The attacker first fills a monitored
cache set with its own lines (Prime), allows the victim to execute, and then
reloads those lines while measuring latency (Probe).  Slow probes indicate
that the victim probably evicted lines from the set.  The observation is at
cache-set granularity rather than at an exact virtual address, and scheduler
activity, prefetching, and unrelated memory traffic introduce noise; refined
contention techniques can improve temporal precision while controlling the
observer effect~\cite{purnal2021primescope}.  By
contrast, Flush+Reload offers line-level resolution but normally relies on
shared physical pages~\cite{Yarom14}.  We select Prime+Probe because the model,
compiled circuit, witness, and proving artifacts need not be mapped into the
attacker's address space.

Prior cache attacks recover DNN architectures from the matrix kernels of native
inference or from compiled executables, learning noise-tolerant trace
representations and matching them against a database of profiled
operators~\cite{cachetelepathy2020,deepcache2024}.  A ZK prover produces a different
trace. The dummy-layout sizing pass, the interleaved synthesis-composite
assignment, and the later polynomial stage each contribute their own structure,
and none of them is a native operator kernel.
Section~\ref{sec:leakage-analysis} analyzes what this trace leaks, and
Section~\ref{sec:attack-design} builds the recovery attack.

\subsection{Threat Model and Recovery Targets}
\label{sec:background-model-information}

\parh{Model as a graph.}
We represent a model architecture as an attributed directed acyclic graph
$G=(V,E)$.  Each node $v\in V$ has an operator type $t_v$ and hyperparameters
$h_v$; for a convolution, $h_v$ gives input/output channels, kernel
dimensions, stride, padding, and groups, and for pooling it gives the window,
stride, and padding.  All evaluated convolutions use unit dilation, so dilation
is fixed rather than recovered or scored.  Edges $E$ specify tensor flow,
including branches, shortcuts, and merge points.

\parh{Recovery goals.}
ZKCache has three recovery goals. Operator recovery estimates the
types $t_v$; hyperparameter recovery estimates the attributes $h_v$; and graph
recovery estimates the edges $E$ on top of the node attributes.  For a sequential
architecture such as VGG~\cite{simonyan2014very}, node order already fixes most
edges, so graph recovery adds little.  For a residual architecture such as
ResNet~\cite{he2016deep}, the same operator sequence can have different identity
or projection shortcuts, so recovering $E$ is a genuinely separate task, and the
one on which we report exact-topology results.

\parh{Scope and deployment.}
We do not claim to recover numerical weights or private inputs; those need different observations and lie outside our scope.  Our threat
model is a hosted proving service that treats the graph and its hyperparameters
as proprietary and keeps the model, the compiled circuit, the proving key, and the witness files on
the server.  The adversary is an unprivileged process co-resident on that server,
without shared pages or privileged counters.  Under this model, recovering
$(t_v,h_v,E)$ is a breach even though the proof transcript stays zero knowledge.

	\section{Architecture Leakage in EZKL}
\label{sec:leakage-analysis}

Section~\ref{sec:background-halo2} established that a Halo2 prover translates
model operators into memory workloads and that EZKL executes the model twice
during proof generation. The first execution sizes the circuit and yields the
\textit{dummy-layout view}; the second synthesizes the proof and yields the
\textit{synthesis-composite view}. We next determine how private architectural
attributes, such as operator type or a convolution's kernel size, filter count,
and stride, govern these memory-access patterns and whether the dependency
persists through cache-set contention. This distinction separates architectural
leakage from an incidental implementation artifact. If the prover's memory-access
patterns are deterministically tied to the model topology, a co-resident
adversary can reconstruct that topology from measured cache contention.

We derive from EZKL's open-source implementation how a model's shape controls
the prover's memory work. These source-derived relations locate the model-dependent signal in
a noisy trace and provide checkable constraints that reject unsupported
configurations. We first define the trace abstraction and methodology
(Section~\ref{sec:source-analysis}). We then compare the two views for operator
types (Section~\ref{sec:operator-type-leakage}), derive convolution relations
(Section~\ref{sec:conv-hyperparameter-leakage}), and state the analysis's scope
and design implications (Section~\ref{sec:scope-challenges}).

\subsection{Methodology and Trace Abstraction}
\label{sec:source-analysis}

Our procedure for finding a leak is the same for every operator.  We keep a
candidate leak only when a private attribute controls a tensor access or a cell
assignment in EZKL's layout code, so that changing the attribute changes the
prover's memory work.  Reading the code establishes that this dependence exists;
Section~\ref{sec:evaluation} tests the separate question of whether it survives
cache-set aggregation and stays measurable.  To label shadow traces for training
and evaluation, we add markers at operator boundaries; these markers are an
offline aid, never available to the adversary, whose only signal is Prime+Probe
contention.

A Prime+Probe measurement is a matrix $X\in\mathbb{R}^{T\times m}$, where $T$ is
the number of probe rounds, $m$ is the number of monitored cache sets, and
$X_{\tau,\sigma}$ is the probe latency for round $\tau$ and monitored set
$\sigma$.  Within this
matrix, each operator segment carries two recurring structures.

\begin{itemize}[leftmargin=*,topsep=2pt,itemsep=1pt]
  \item a \emph{unit signature} that captures the cache sets, duration, and burst pattern of
  one repeated unit of work, such as one dot-product layout, one pooling-window
  sort, or one elementwise gate pass; and
  \item a \emph{repetition profile} that captures how often that unit recurs, and whether the
  recurrences form a second-level spatial or channel period.
\end{itemize}

Both structures are fixed by the model's shape.  An operator's loop bounds and
tensor dimensions set its unit signature and repetition profile, whereas the
numeric values in the weights and inputs do not; ZKCache therefore recovers a
model's architecture, not its weights.  The rest of this section derives, view by
view, how operator type and convolution parameters control these two structures.
How ZKCache encodes the resulting segments is deferred to
Section~\ref{sec:attack-design}.

\subsection{Operator-Type Signatures Across the Two Views}
\label{sec:operator-type-leakage}

For exposition, we first characterize the evidence that separates operators
and then examine operator-specific parameters.  This order follows the
prover, as the operator type determines which layout routine EZKL
dispatches, and the hyperparameters then set that routine's loop bounds and
access extents.  It does not impose a type-first decoder.  ZKCache retains the
type $t$ and hyperparameters $h$ as a joint key $q=(t,h)$ and resolves them
together during decoding (Section~\ref{sec:attack-design}).  In this white-box
analysis we examine one operator's source path at a time.  No victim-time
boundary is assumed; Section~\ref{sec:attack-design} explains how ZKCache
estimates those boundaries from an unlabeled trace.

\begin{figure}[t]
  \centering
  \includegraphics[width=\columnwidth]{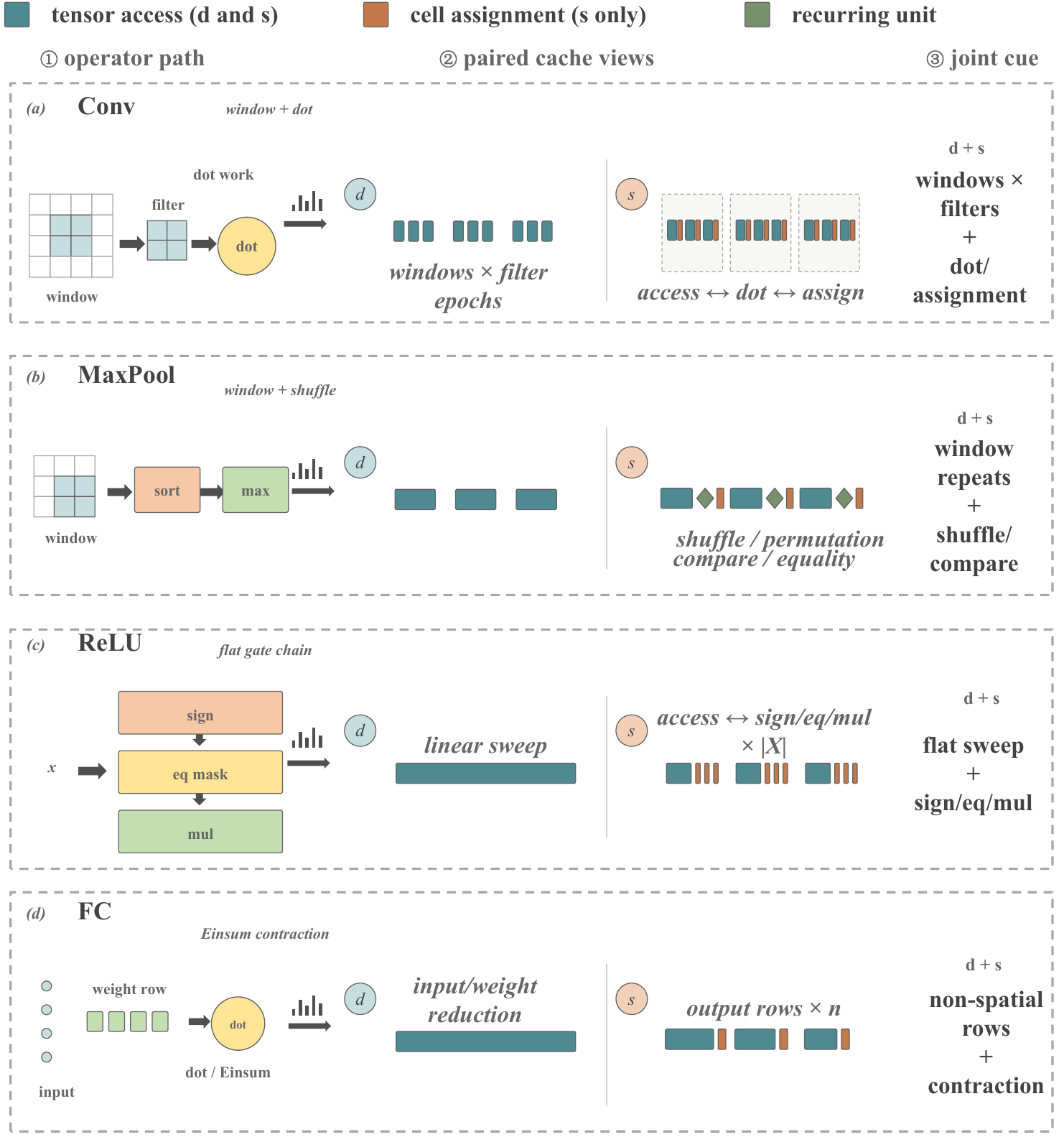}
  \caption{Source-derived two-view signatures for four representative
  operators.  The dummy-layout view highlights traversal and repetition; the
  synthesis-composite view adds interleaved tensor-access and cell-assignment
  activity.  The two views jointly distinguish operator types that can be
  ambiguous in either view alone.}
  \label{fig:operator-traces}
\end{figure}

\parh{Observation 1. Operator types induce complementary signatures across the
two temporal views.}
EZKL dispatches every graph node to an operator-specific layout.  In the
dummy-layout view $d$, that layout exposes its traversal topology, namely which tensors
it visits, how its loops nest, and whether it walks windows, filters, or flat
vectors.  The synthesis-composite view $s$ preserves that traversal and adds the
gadget execution and cell writes on top of it.  Because tensor access and cell
assignment stay interleaved on the single $s$ timeline, they act as two distinct
cues within one view rather than as extra phases. Together, their unit signatures
and repetition profiles separate operators that collide in either view. Both the
traversal topology visible in
$d$ and the gadget composition added in $s$ follow from the layout routine EZKL
dispatches for each operator type, so we derive them by reading that routine.
Figure~\ref{fig:operator-traces} sketches the paired signatures of four
representative operators; the detailed comparison appears in
Appendix~\ref{app:operator-signatures}
(Table~\ref{tab:operator-signatures}).

We organize the comparison into windowed operators (Conv and MaxPool) and
non-windowed operators (ReLU and FC), comparing first within each group and then
across them.

\parh{Convolution and pooling.}
Convolution and pooling both slide a window across the input and materialize one
window per output coordinate, so in the dummy-layout view their repeated-window
traversals look alike.  Convolution additionally sweeps a filter dimension,
nesting each spatial sweep inside a filter epoch.  The synthesis-composite view
then separates the two.  Convolution interleaves image and filter access with
dot-product gadgets, an optional bias, and cell assignment, whereas pooling reads
no filter and instead interleaves window access with sorting, permutation,
comparison, and equality work.  The composite view therefore resolves the
ambiguity that window repetition alone leaves.

\parh{Activation and fully connected layers.}
ReLU and FC do not construct a sliding spatial window.  In the dummy-layout
view, ReLU makes a few flat, elementwise passes, whereas FC performs repeated,
non-spatial input-and-weight reductions.  In the synthesis-composite view, ReLU
adds sign decomposition, equality masking, and multiplication, whereas FC adds
matrix-contraction assignments.  Their repetition depth and gadget composition
therefore separate them.

\parh{Convolution and fully connected layers.}
Convolution and FC collide in the reverse direction from convolution and
pooling.  Both reduce vectors, so both produce dot-like gadget and assignment
activity in the synthesis-composite view.  The dummy-layout view supplies the
distinction, because convolution shifts a local window and revisits it across
filter epochs, whereas the FC traversal has no spatial extent to shift.  This
case establishes the second direction of Observation 1, in which $d$ resolves
what $s$ leaves ambiguous.

No view-independent signature identifies every operator, and the two views fail
in different directions. ZKCache therefore encodes each view separately. Within
a view, it forms operator-type
evidence by minimizing over the profiled hyperparameter keys of that type, while
the candidates it carries forward remain joint keys $q=(t,h)$.
Section~\ref{sec:attack-design} formalizes this retrieval and fusion.

\subsection{Two-View Convolution Hyperparameter Leakage}
\label{sec:conv-hyperparameter-leakage}

\subsubsection{Convolution Shape and a Running Example}
\label{sec:conv-running-example}

ZKCache evaluates all preregistered fields of supported parameterized operators,
the per-operator field sets that we fix in advance and score in
Section~\ref{sec:evaluation}.  The detailed source analysis here focuses on
convolution, whose rich parameterization exposes explicit links between tensor
shape and memory activity. These links yield additional analytical constraints.
Non-convolution parameters rely on profiled joint-key retrieval and generic shape
constraints (Section~\ref{sec:attack-design}). We illustrate the convolution
relations with a running example. EZKL instantiates the same layout over batch,
groups, channels, and spatial dimensions, so the relations below hold for any
convolution. We return to the other operators at the end of this subsection. The
running example has
input shape
$1\times3\times224\times224$, uses 64 filters of shape $3\times3\times3$ with
stride one and one-cell padding, and produces output of shape
$1\times64\times224\times224$.

For a two-dimensional grouped convolution with unit dilation, let $B$ be the
batch size; $C_{in}$ and $C_{out}$ the input and output channels;
$H_{in}\times W_{in}$ and $H_o\times W_o$ the input and output spatial sizes;
$K_h\times K_w$ the kernel size; $s_h,s_w$ the strides; $p_t,p_b,p_l,p_r$ the
top, bottom, left, and right padding; and $g$ the number of groups.  EZKL
computes
\begin{equation}
\begin{aligned}
 H_o&=\left\lfloor\frac{H_{in}+p_t+p_b-K_h}{s_h}\right\rfloor+1,\\
 W_o&=\left\lfloor\frac{W_{in}+p_l+p_r-K_w}{s_w}\right\rfloor+1.
\end{aligned}
\label{eq:conv-output}
\end{equation}
The dot-product length $L_{dot}$ and number of dot products $N_{dot}$ are
\begin{equation}
 L_{dot}=\frac{C_{in}}{g}K_hK_w, \qquad
 N_{dot}=B C_{out}H_oW_o.
\label{eq:conv-observables}
\end{equation}
For the running example, $L_{dot}=27$.  Each filter performs $224\times224$ dot
products, and all 64 filters perform 3,211,264 dot products. These are
source-derived properties of the trace for this known configuration, not
quantities Prime+Probe reads from a single trace. The convolution layout places
the output channel outside the spatial
coordinates, so the trace is organized as 64 filter epochs, each a
$224\times224$ spatial sweep of 27-element image-and-kernel contractions.

Appendix~\ref{app:conv-source} reproduces the relevant layout statements in
Listing~\ref{lst:conv-source}.  For each output element, they set the window
start to the output coordinate times the stride, take the window extent from
the kernel dimensions, and hold the filter index fixed across one spatial
sweep.  They then copy the selected image window and filter slice into fresh
tensors before the dot product.  The same statements run in both views, which
is why the two views share this hierarchy.  In the dummy-layout view they only
read the image and kernel footprints, while in the synthesis-composite view the
identical reads are interleaved with the cell assignments they feed.

\subsubsection{Evidence Across the Two Views}

With the running example, we examine three evidence streams.  The first is the
dummy-layout stream, carried by the dummy-layout view $d$.  The other two, the
tensor-access stream and the assignment stream, are interleaved within the
single synthesis-composite view $s$. The streams retain different information.
The dummy-layout stream records the amount of work and repetition but not a
stable window order. The tensor-access stream restores that order and the stride
evidence that depends on it. The assignment stream adds the write volume.

\parh{Dummy-layout stream.}
\label{sec:witness-leakage}
The dummy-layout stream reflects the work performed per output element and the
layer's total repetition.  During circuit sizing, EZKL runs the same convolution
layout without a backing Halo2 region, so cell assignment is switched off.  In
place of writing a Halo2 cell, an assignment call advances the running row and cell
counters that determine the circuit's size and returns a placeholder value, so
the tensor traversal and the gadget computation are unchanged
(Section~\ref{sec:background-systems}).  The sizing pass therefore still pads the
image, builds each window and filter slice, flattens them, and computes every dot
product.  For each output element, the image and filter slices contain $L_{dot}$
entries.  A larger kernel therefore expands the slice-copy footprint and the
per-element duration.  For fixed batch and output-grid dimensions, the layout
repeats this work for all $N_{dot}$ output elements, so increasing $C_{out}$
adds full spatial sweeps rather than enlarging the work within one window.
Stride shifts successive window starts and, through
Equation~\ref{eq:conv-output}, changes the number of windows.  In a cache trace
these effects appear as changes in the segment's unit-signature distribution and
repetition profile, not as exact element or window counts.

Because the sizing pass maps over output elements in parallel, it can reorder
nearby windows, so this stream exposes the total repetition and the number of
filter epochs, but not a reliable left-to-right window order.  Kernel size and
filter count are therefore stronger signals here than stride, which depends on
that order. A
high-performance inference library would instead fuse the layer into blocked
matrix-multiplication or Winograd kernels~\cite{chen2018tvm,lavin2016fast},
blurring the mapping from model dimensions to memory accesses; EZKL leaves that
mapping explicit.

\parh{Tensor-access stream.}
\label{sec:proof-tensor-reads}
Proving enumerates the output elements sequentially rather than in parallel, so
it recovers the stable window order the sizing pass loses, and with it the
firmest hierarchy of the three streams.  It runs the same convolution layout a
second time inside Halo2 with a real region, a contiguous block of circuit rows;
because the intermediate activations were not saved, the image and kernel
accesses repeat (Section~\ref{sec:background-systems}).  The outer order of
windows and filter epochs therefore follows the Cartesian iteration order, giving

\begin{equation}
\underbrace{\big[\text{window unit}\big]^{H_oW_o}}_{\text{one filter epoch}}
\quad\text{repeated}\quad C_{out}\ \text{times}.
\label{eq:filter-epochs}
\end{equation}

This hierarchy supplies three cues.  First, the duration and cache-set footprint
of one window stay tied to $L_{dot}$.  Second, a kernel-slice change after every
$H_oW_o$ windows marks a candidate filter boundary, which exposes $C_{out}$
independently of total runtime.  Third, adjacent image windows start $s_w$
elements apart horizontally and $s_hW_{pad}$ elements apart vertically, where
$W_{pad}=W_{in}+p_l+p_r$ is the padded input width, so their cache-line overlap
changes with stride.  At cache-set granularity this displacement is approximate,
so we combine it with the inferred output-grid size.

\parh{Assignment stream.}
\label{sec:advice-write-leakage}
The assignment stream adds the write volume of the layer to the same timeline,
reinforcing the kernel and filter evidence while adding little on stride.  Only
the real proving run assigns cells.  For each output element EZKL reads the slices,
runs the dot-product gadget, and writes the resulting cells before moving on, so
reads and writes alternate at fine granularity.  Each dot product pads its two
operand vectors to the column width $\omega$, writes the operands and
accumulators, enables the gadget's selectors, and advances the region by the
assigned length; an optional bias adds one more gadget.
Figure~\ref{fig:advice-layout} shows this assignment for one output element.

\begin{figure}[t]
\centering
\begingroup
\setlength{\fboxsep}{3pt}
\fbox{\begin{minipage}{\dimexpr\columnwidth-2\fboxsep-2\fboxrule\relax}
\scriptsize\ttfamily
\setlength{\tabcolsep}{4pt}
\begin{tabular*}{\linewidth}{@{\extracolsep{\fill}}llll@{}}
row & input-0 & input-1 & output / selector \\
\midrule
0 & kernel preload & image preload & -- \\
$r$ & $w_0$ & $x_0$ & $acc_0$ / DotInit \\
$r+1$ & $w_1$ & $x_1$ & $acc_1$ / Dot \\
$\cdots$ & $\cdots$ & $\cdots$ & $\cdots$ \\
$r+L-1$ & $w_{L-1}$ & $x_{L-1}$ & $acc_{L-1}$ / Dot \\
$r+L$ & bias & $acc_{L-1}$ & output / Add \\
\end{tabular*}
\end{minipage}}
\endgroup
\caption{Cell-assignment slice of one convolution output within the
synthesis-composite timeline.  Here $r$ is the first
dot row, $L$ is the assigned dot span, and $w_i$, $x_i$, and $acc_i$ are the
kernel, input, and accumulator values at step $i$.  Halo2 stores columns as
in-memory vectors; duplication at column boundaries is omitted.}
\label{fig:advice-layout}
\end{figure}

The attacker learns neither the cell values nor a literal rectangular array; its
Prime+Probe measurements reflect contention from the backing column vectors,
selector state, and copy metadata.  These accesses are regular because a cell's
column and row coordinates are derived from the region's linear position.

Writing $A_{conv}$ for the total operand-and-accumulator assignment volume of the
convolution, and ignoring the initial tensor assignment, bias work, and
configuration-dependent boundary constants,
\begin{equation}
 A_{conv}=\Theta\!\left(
 N_{dot}\left\lceil L_{dot}/\omega\right\rceil \omega\right),
\label{eq:advice-volume}
\end{equation}
where $\lceil L_{dot}/\omega\rceil\,\omega$ is the padded length of one dot
product and $N_{dot}$ its repetition count.  Kernel size therefore sets the width
of each assignment block, while filter count and the output grid set how many
blocks occur.

Prime+Probe captures only a noisy projection of these writes, not their values,
so the block width and repetition register as distributions rather than exact
counts.  Stride does not enter the dot-product gadget at all, and affects
assignment only indirectly through $H_oW_o$ and the block count; the ordered
tensor accesses in the same segment therefore carry the stronger stride
evidence.

\subsubsection{Ambiguities and Joint Constraints}
\label{sec:leakage-to-properties}

\parh{Observation 2. Each convolution parameter is constrained by a different
combination of the three streams.}
Table~\ref{tab:parameter-evidence} records, for each parameter, whether a
stream's effect is \emph{primary}, meaning the parameter sets a loop bound or
access extent in the layout, or \emph{secondary}, meaning it acts only through a
derived quantity such as the output-grid size.  A primary effect still does not
let the parameter be read from a single cache set.  Kernel size and filter count
leave primary effects in all three streams, whereas stride is primary only in
the ordered tensor-access stream, so no stream recovers every parameter alone
and the streams must be combined.  We take the three parameters in turn.

\begin{table*}[t]
\centering
\small
\caption{How each convolution parameter affects the three evidence streams.
Primary effects set a loop bound or access extent; secondary effects act through
a derived quantity such as the output-grid size.}
\label{tab:parameter-evidence}
\begin{tabular}{@{}p{0.19\textwidth}p{0.24\textwidth}p{0.24\textwidth}p{0.25\textwidth}@{}}
\toprule
Evidence stream & Kernel-size evidence & Filter-count evidence & Stride evidence \\
\midrule
Dummy-layout stream & Primary effect on image/filter slice length and dot work;
ordering may be parallel & Primary effect on output-element count and repeated
filter epochs & Secondary effect through output count and aggregate overlap;
fine order is weak \\
Tensor-access stream & Primary effect on per-window footprint and dot work &
Primary effect on ordered $H_oW_o$ epochs and kernel-slice switches & Primary
effect on ordered window displacement; also changes $H_o,W_o$ \\
Assignment stream & Primary effect on cells and duplicated blocks per dot &
Primary effect, jointly with the spatial grid, on total dot blocks & Secondary
effect through output-grid size and assignment count \\
\bottomrule
\end{tabular}
\end{table*}

\parh{Kernel size.}
Kernel size scales the per-window slice footprint and dot work, and thus the
unit signature, through $L_{dot}$ (Equation~\ref{eq:conv-observables}).  For the
running example the $3\times3$, $C_{in}=3$, $g=1$ layer gives $L_{dot}=27$,
distinct from the $1\times1$ or $5\times5$ candidates.  The signature is not
unique, however.  In a general network the same $L_{dot}$ can arise from other
combinations of channels, groups, and kernel area, so the attack retains the
plausible $(K_h,K_w,g)$ tuples until the paired views and the neighboring layers
resolve the collision.

\parh{Number of filters.}
Filter count is the most strongly constrained of the three, because two cues
within one synthesis-composite segment corroborate it.  The tensor-access
stream marks a candidate epoch boundary every $H_oW_o$ windows, and the
assignment stream supplies the block repetition that $N_{dot}$ predicts
(Equation~\ref{eq:conv-observables}).  For the running example this is a
$224\times224$ epoch repeated 64 times.  The count is still not read exactly, as
no filter need occupy a unique cache set, and the evidence lies in the segment as
a whole rather than in any single block.

\parh{Stride.}
Stride is the weakest and most ambiguous signal, and it appears two ways.
First, Equation~\ref{eq:conv-output} predicts the output-grid repetition for each
candidate.  Second, the tensor-access stream of a synthesis-composite segment
reflects the displacement and overlap between adjacent windows.  Away from
boundaries, the geometric horizontal and vertical overlap fractions are
\begin{equation}
 \begin{aligned}
 \rho_w(s_w)&=\frac{K_h\max(0,K_w-s_w)}{K_hK_w},\\
 \rho_h(s_h)&=\frac{K_w\max(0,K_h-s_h)}{K_hK_w},
 \end{aligned}
\label{eq:window-overlap}
\end{equation}
where $\rho_w$ and $\rho_h$ are the fractions of input elements shared between
adjacent horizontal and vertical windows.  Cache lines, parallel copying,
prefetching, and set conflicts distort these fractions. Equation~\ref{eq:window-overlap} is a
consistency relation, not a direct stride
decoder.  A candidate stride must agree at once with the output repetition, the
horizontal and vertical overlap, and the assignment-side block span, and because
padding can absorb a stride change, several stride-and-padding tuples may
survive.

\parh{Joint consistency.}
These per-parameter ambiguities need not resolve into a unique architecture by
themselves. A long unit signature cannot distinguish a larger kernel from more
input channels. A large assignment footprint cannot distinguish more filters
from a larger output grid. A candidate admitted by one stream is discarded when
another leaves it unsupported.
Equations~\ref{eq:conv-output}, \ref{eq:conv-observables},
\ref{eq:window-overlap}, and \ref{eq:advice-volume} reject tuples whose loop and
assignment relations are inconsistent, and adjacent layers further constrain each
other because their tensor shapes must match. The attack does not fix an
operator's type or hyperparameters locally. It carries several joint keys per
operator and resolves them jointly in
Section~\ref{sec:attack-design}.

\parh{Beyond convolution.}
The derivation generalizes to every operator EZKL lowers, because each
operator's layout takes its loop bounds and access extents from that operator's
hyperparameters in the same way, though the resulting relation differs. A fully
connected layer is the specialization of Equation~\ref{eq:conv-observables} with
$g=1$, $K_h=K_w=1$, and $H_o=W_o=1$, giving unit length $L_{dot}=C_{in}$
repeated $N_{dot}=BC_{out}$ times, so its input and output widths play the roles
kernel size and filter count play for convolution.  Max pooling keeps
Equation~\ref{eq:conv-output} for its output grid but has no filter epoch, so
its unit is a sort over the $K_hK_w$ window elements and its repetition is
$BC_{in}H_oW_o$; this is what makes its window and stride recoverable in
Section~\ref{sec:evaluation}.  A ReLU activation has a constant unit and a
repetition equal to its tensor size. We derive convolution in full because it
has the largest hyperparameter space and candidate set. For the other operators,
the unit signature and repetition profile leave a smaller space, which ZKCache
resolves with profiled joint-key
retrieval and generic shape constraints (Section~\ref{sec:attack-design}).

\subsection{Scope and Design Challenges}
\label{sec:scope-challenges}

This subsection bounds what the two views do \emph{not} reveal, checks that the
analysis is not specific to EZKL, and derives design requirements for the attack.

\parh{Cryptographic proving.}
After synthesis, Halo2 commits to the columns and builds the permutation, lookup,
and quotient arguments~\cite{halo2book}.  This stage is dominated by whole-circuit
polynomial operations, namely fast Fourier transforms and multi-scalar
multiplications, so its footprint tracks the total circuit size rather than
individual layers.  A co-resident process can therefore still bound that total
size and locate where synthesis ends, but cannot label a specific layer or
stride, because a committed column mixes many operators and no longer preserves a
per-layer timeline.

\parh{TeleSparse.}
TeleSparse applies sparsification and neural teleportation~\cite{telesparse2025}
but keeps the same PLONKish circuit structure and still runs the model twice, so
both the dummy-layout and synthesis-composite views still apply.  Only its
concrete signatures change, because its compiled binary and operator code differ
from EZKL's.  Section~\ref{sec:evaluation} tests whether re-profiling on
TeleSparse preserves the attack's accuracy.

\parh{From leakage to design requirements.}
The analysis sharpens the three challenges introduced in
Section~\ref{sec:introduction} into concrete requirements.  Phase isolation
becomes extracting and aligning the dummy-layout and synthesis-composite segments
of each operator under scheduling noise.  Shape-to-memory attribution becomes
preserving one synthesis-composite sequence, separating its tensor-access and
assignment streams, and keeping several joint keys per operator.
Structure beyond a layer sequence becomes enforcing graph-wide shape and
residual-edge consistency.  These requirements drive the attack design in
Section~\ref{sec:attack-design}.

	\section{ZKCache Design and Implementation}
\label{sec:attack-design}

\begin{figure*}[t]
  \centering
  \includegraphics[width=\textwidth]{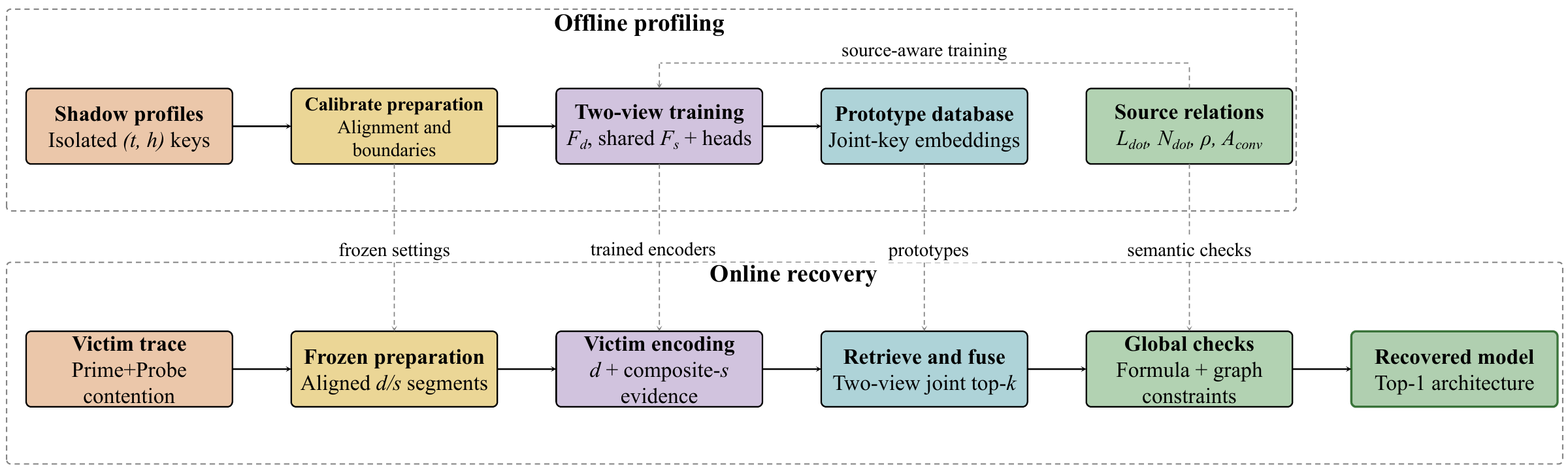}
  \caption{ZKCache workflow.  Offline profiling fixes preparation, learns the
  two-view representation, and builds joint-key prototypes.  The online path
  applies these frozen components to a victim trace, then combines local evidence
  with analytical and graph constraints to recover the architecture.}
  \label{fig:attack-overview}
\end{figure*}

The challenges in Sections~\ref{sec:introduction} and
\ref{sec:scope-challenges} require more than one measurement. A cache trace does
not mark operator boundaries, no view separates every operator type, and a local
decision about one segment is unreliable. ZKCache retains weak evidence until
later stages can evaluate it. Trace preparation isolates and aligns the two views
of each operator for phase isolation. Separate encoders and two
synthesis-composite pooling heads address shape-to-memory attribution. Retrieval
and fusion retain several joint keys per operator. Constrained decoding evaluates
them against the analytical constraints and the network's shape to recover
structure beyond a layer sequence. Figure~\ref{fig:attack-overview} summarizes
the workflow.

\subsection{Design Overview}
\label{sec:design-overview}

ZKCache preserves the prover's execution structure, and each stage leaves later
stages able to revise its decisions. A dedicated encoder $F_d$ and a shared
encoder $F_s$, one neural network per view, model the different schedules of the
dummy-layout view $d$ and the synthesis-composite view $s$. Two pooling heads
over $F_s$ keep the tensor-access and assignment cues separable without splitting
the single $s$ timeline into extra views.

\subsection{Shadow Profiling and Trace Preparation}
\label{sec:shadow-dataset}

A proof request yields one undifferentiated latency matrix. Nothing in it marks
where the dummy-layout view ends, where the synthesis-composite view begins, or
where each operator's segment lies inside either, so trace preparation must
recover those boundaries from the trace alone, using detectors fixed beforehand
on shadow data.

\parh{Target systems.}
We target EZKL 23.0.5 and TeleSparse 1.0.0 under native CPU proving.  VGG16 and
ResNet-18 are imported into both systems with fixed input, quantization, circuit,
and proving configurations.  In EZKL, view $d$ covers \texttt{gen\_witness},
\texttt{GraphCircuit::forward}, and \texttt{Model::dummy\_layout}, and view $s$
begins when \texttt{create\_proof} invokes \texttt{GraphCircuit::synthesize}.
Within $s$, tensor traversal and \texttt{RegionCtx} writes alternate per output
element.  TeleSparse exposes the corresponding dummy-layout and
synthesis-composite intervals.

\parh{Shadow profiles.}
Before measuring any victim request, the adversary compiles and runs models of
its own on the same system and hardware, drawing operator configurations from
image-classification models in the Open Neural Network Exchange (ONNX) Model
Zoo~\cite{onnxmodelzoo}, so the profiled configurations are the ones deployed
classifiers use.  VGG16 and ResNet-18 are excluded as source models.  After ONNX
shape inference, each supported node is normalized into an operator type $t$ and
a complete parameter key $h\in\mathcal K_t$, and duplicate keys are removed.  For
convolution,
\begin{equation}
 h=(B,C_{in},H_{in},W_{in},C_{out},K_h,K_w,s_h,s_w,p,g),
 \label{eq:conv-key}
\end{equation}
where $p=(p_t,p_b,p_l,p_r)$.  Each key is compiled as an isolated single-operator
circuit rather than inside a whole shadow network, so its segment carries that
key alone.  A key the corpus never produced then has no prototype, which makes
profiled-key coverage (Table~\ref{tab:target-statistics}) the figure of merit for
the database.

\parh{Shadow records.}
For every joint key $q=(t,h)$, ZKCache compiles an isolated circuit for each
target system and collects 60 complete traces across fresh inputs, process
restarts, allocations, and scheduler conditions.  A raw shadow record is
\begin{equation}
 (X_{q,j},q,c_j,\ell_{q,j}),
 \label{eq:shadow-record}
\end{equation}
where $X_{q,j}$ is the latency matrix of Section~\ref{sec:source-analysis} for
execution $j$, $c_j$ stores system, hardware, and collection metadata, and
$\ell_{q,j}$ contains view and operator boundaries obtained from a separate
instrumented execution.  We partition the source graphs before collecting any
trace, so that no two segments from one execution can land in different splits;
Section~\ref{sec:experimental-setup} lists what each split fixes.

\parh{Trace preparation.}
The collector constructs eviction sets for 2,048 candidate last-level-cache sets,
primes them, waits 2,000 cycles, and probes in permuted order.  Mutual
information selects $m=64$ informative, nonredundant sets per system and hardware
profile, so one sample is a latency vector $x_\tau\in\mathbb R^m$.  ZKCache
standardizes each set by its calibration median and median absolute deviation,
suppresses isolated events, and preserves both latency and set identity.  A
request's start and end are visible to a co-resident process as the onset and
cessation of sustained prover activity, and within that window normalized
cross-correlation against complete-request profiles locates the coarse
dummy-layout, synthesis-composite, and cryptographic-proving intervals.  Dynamic
time warping then refines alignment within a view~\cite{sakoe1978dtw}.

\parh{Operator boundaries.}
Within each view, change-point detection proposes operator boundaries from
activity, footprint, and burst structure, and a dynamic program chooses a
monotone boundary sequence using duration ranges and boundary priors frozen on
shadow validation~\cite{killick2012changepoint}.  The two views share operator
order but may have different boundaries.  A view whose alignment or boundary
confidence falls below the threshold fixed on calibration is dropped at that
position rather than imputed, and Equation~\ref{eq:multi-stage-fusion}
renormalizes over the views that remain.  The output at position $i$ is therefore
a set of aligned segments $S_i^{(v)}$ indexed by the available views
$\mathcal V_i\subseteq\{d,s\}$.  Instrumented
shadow runs supply the segmentation labels and the oracle-segmentation upper
bound, in which ground-truth boundaries replace the detector; victim traces use
only the frozen detector.

\subsection{Prover-Aware Two-View Representation}
\label{sec:two-view-representation}

\parh{One encoder per view, two pooling heads over $s$.}
Because the two views resolve different operator ambiguities
(Section~\ref{sec:operator-type-leakage}), one representation cannot carry the
evidence of both. ZKCache encodes each view with
its own encoder, $F_d$ for the dummy-layout view and $F_s$ for the
synthesis-composite view.  Within the synthesis-composite view, tensor access and
cell assignment remain interleaved on one timeline, so they are two cues within
one sequence, not two sequences. $F_s$
therefore consumes the complete synthesis-composite segment once and is shared by
the tensor-access and assignment pooling heads $H_{\mathrm{ta}}$ and
$H_{\mathrm{as}}$, which Table~\ref{tab:stage-results} measures against the same
two heads on untied encoders.  Each segment is resampled to a fixed temporal
resolution and normalized with training statistics, and the embeddings are
\begin{equation}
 \begin{aligned}
 z_i^{(d)}&=\operatorname{norm}\bigl(F_d(S_i^{(d)})\bigr),\\
 u_i^{(s)}&=F_s(S_i^{(s)}),\qquad
 z_i^{(s,b)}=\operatorname{norm}\bigl(H_b(u_i^{(s)})\bigr),\\
 &\hspace{35mm}b\in\{\mathrm{ta},\mathrm{as}\}.
 \end{aligned}
 \label{eq:view-embedding}
\end{equation}
We write $\mathcal B=\{d,(s,\mathrm{ta}),(s,\mathrm{as})\}$ for the three learned
representations, one per evidence stream of
Section~\ref{sec:conv-hyperparameter-leakage}.  Both encoders contain three
residual one-dimensional convolution blocks with 64, 128, and 128 channels.
$F_d$ uses global average pooling, whereas $F_s$ retains its feature sequence for
the two attention-pooling heads.  Each head produces a 128-dimensional normalized
embedding, and a 64-dimensional projection maps into the shared two-view space.

\parh{Contrastive objective within a representation.}
Within one representation $c\in\mathcal B$, independent executions of the same
joint key are positives and traces of different keys are negatives, and
configurations whose source descriptors are close
(Equation~\ref{eq:source-distance}) are sampled preferentially as hard negatives.
ZKCache trains each representation with the supervised contrastive loss
$\mathcal L_{intra}$ of Khosla et al.~\cite{khosla2020supervised}, computed at
temperature $\tau$ over cosine similarities between the normalized embeddings.
Augmentation is limited to 10\% time warp, 0.05-standard-deviation latency
jitter, short gaps covering at most 1\% of a segment, and 10\% monitored-set
dropout, which perturb timing and set occupancy without changing the unit
signature or repetition profile that Section~\ref{sec:source-analysis} identified
as the carriers of shape.  We train for 120 epochs with AdamW at batch size 256,
learning rate $3\times10^{-4}$, weight decay $10^{-4}$, and $\tau=0.07$.

\parh{Aligning the two views.}
The two synthesis-composite embeddings are combined into one,
$z_i^{(s)}=\operatorname{norm}(W_c[z_i^{(s,\mathrm{ta})};z_i^{(s,\mathrm{as})}])$
with $W_c$ learned on shadow data.  View projections $G_d$ and $G_s$ map
$z_i^{(d)}$ and $z_i^{(s)}$ into a shared space, in which the cross-view loss
$\mathcal L_{cross}=\sum_{q,j}\lVert G_d(z_{q,j}^{(d)})-G_s(z_{q,j}^{(s)})
\rVert_2^2$ pulls together the two views of one shadow execution.

\parh{Source-derived supervision.}
The evidence streams constrain different parameters. Stride is primary only in
the tensor-access stream, while assignment volume is a kernel-and-filter cue.
ZKCache gives each convolution key one source descriptor per representation
rather than one shared descriptor. Writing
$\ell_L=\log(1+L_{dot})$, $\ell_N=\log(1+N_{dot})$, and
$\ell_A=\log(1+A_{conv})$,
\begin{equation}
 \begin{aligned}
 \psi^{(d)}(h)=\psi^{(s,\mathrm{ta})}(h)
   &=\bigl[\ell_L,\ell_N,\rho_w,\rho_h\bigr],\\
 \psi^{(s,\mathrm{as})}(h)
   &=\bigl[\ell_L,\ell_N,\ell_A\bigr],
 \end{aligned}
 \label{eq:source-semantic-vector}
\end{equation}
so the dummy-layout representation is not supervised by assignment volume, and
the assignment representation is not forced to encode window displacement.  For
$c\in\mathcal B$, the corresponding source distance is
\begin{equation}
 d_{src}^{(c)}(h_i,h_j)=
 \bigl\|\Lambda_c\bigl(\psi^{(c)}(h_i)-\psi^{(c)}(h_j)\bigr)\bigr\|_1,
 \label{eq:source-distance}
\end{equation}
where the diagonal $\Lambda_c$ standardizes each descriptor on the training
split.  We cap it as $\bar d_{src}^{(c)}=\min(d_{src}^{(c)}/\kappa_c,1)$, with
$\kappa_c$ fixed to the 95th percentile of training-pair distances.  Source
projections $\widetilde G_c$, with $\widetilde G_d=G_d$ and one linear projection
per pooling head, carry the available convolution pairs
$\mathcal P_{conv}^{(c)}$ into the space in which
\begin{equation}
 \mathcal L_{src}=\sum_{c\in\mathcal B}
 \sum_{(i,j)\in\mathcal P_{conv}^{(c)}}
 \Bigl|\bigl\|\widetilde G_c(z_i^{(c)})-\widetilde G_c(z_j^{(c)})\bigr\|_2-
       \bar d_{src}^{(c)}(h_i,h_j)\Bigr|
 \label{eq:source-aware-loss}
\end{equation}
trains the distance between two convolution embeddings to track the
source-derived distance between their configurations. Training minimizes
$\mathcal L_{rep}=\mathcal L_{intra}+\alpha\mathcal L_{cross}
+\beta\mathcal L_{src}$. The first two terms make the embedding tolerant of
measurement noise, and the third defines proximity between convolution
configurations (Table~\ref{tab:stage-results}).
Section~\ref{sec:experimental-setup} lists the held-out splits that fix $\alpha$,
$\beta$, and the decode weights below.

\parh{Prototype database.}
After training, every segmented shadow run enters a system- and
hardware-specific database.  For each key $q$ and representation
$c\in\mathcal B$, the database stores the individual embeddings and their class
prototype~\cite{snell2017prototypical},
\begin{equation}
 p_c(q)=\operatorname{norm}\Bigl(|\mathcal I_c(q)|^{-1}
 \sum_{j\in\mathcal I_c(q)}z^{(c)}_{q,j}\Bigr),
 \label{eq:embedding-prototype}
\end{equation}
where $\mathcal I_c(q)$ indexes the training runs of $q$ for which
representation $c$ is available.  The two synthesis-composite prototypes come
from the same segment and the same $F_s$ and differ only in the pooling head.

\subsection{Candidate Retrieval and Evidence Fusion}
\label{sec:retrieval-fusion}

\parh{Two-view distances.}
For every available victim embedding, ZKCache compares it with the database
prototypes.  The dummy-layout distance and the two synthesis-composite distances
of a joint key $q=(t,h)$ are cosine distances,
\begin{equation}
 \begin{aligned}
 d_d(q\mid i)&=1-\operatorname{sim}(z_i^{(d)},p_d(q)),\\
 d_{s,b}(q\mid i)&=1-\operatorname{sim}(z_i^{(s,b)},p_{s,b}(q)),\\
 d_s(q\mid i)&=\sum_{b\in\{\mathrm{ta},\mathrm{as}\}}\eta_{ib}d_{s,b}(q\mid i),
 \quad \textstyle\sum_b\eta_{ib}=1,
 \end{aligned}
 \label{eq:embedding-distance}
\end{equation}
where the pooling-head reliabilities $\eta_{ib}$ are frozen on calibration data.
The two heads therefore contribute two pieces of evidence about one segment and
do not create separate candidates or segments.  We use exhaustive cosine search
and combine the prototype distance with the mean distance to the five nearest
shadow runs, of which Equation~\ref{eq:embedding-distance} gives the prototype
component.  Operator evidence is obtained by minimizing over the profiled
hyperparameter keys of that type,
\begin{equation}
 d_v^{op}(t\mid i)=\min_{h\in\mathcal K_t}d_v((t,h)\mid i),
 \qquad v\in\{d,s\},
 \label{eq:operator-embedding-distance}
\end{equation}
but ZKCache ranks all profiled joint keys together and so never fixes a type
before its hyperparameters, as Section~\ref{sec:operator-type-leakage} required.
Convolution keys additionally use the source descriptors above, whereas
non-convolution hyperparameters rely on profiled-key retrieval and generic shape
constraints (Section~\ref{sec:constraint-filtering}).

\parh{Candidate retention.}
Let $\mathcal Q$ be the profiled joint keys the frozen database holds for the
target system and hardware profile.  Each available view ranks them
independently,
\begin{equation}
 \mathcal C_v^{(k)}(i)=
 \operatorname*{TopK}_{q\in\mathcal Q}\bigl[-d_v(q\mid i)\bigr],
 \label{eq:stage-topk}
\end{equation}
and ZKCache retains the union
$\mathcal U_i=\bigcup_{v\in\mathcal V_i}\mathcal C_v^{(k)}(i)$ rather than the
intersection, because the two views fail in different directions
(Section~\ref{sec:operator-type-leakage}) and an intersection would discard a
candidate that one view supports strongly.  Ties within a calibrated margin of
the $k$th entry are also admitted, and the validation split fixes $k$ before any
victim trace is collected.

\parh{Calibrated fusion.}
Cosine distances have different distributions across the two views, and a
held-out calibration split maps them to comparable distances $\widetilde d_v$.
ZKCache then computes
\begin{equation}
 D_{fuse}(q\mid i)=
 \frac{\sum_{v\in\mathcal V_i}w_{iv}\widetilde d_v(q\mid i)}
      {\sum_{v\in\mathcal V_i}w_{iv}},\qquad q\in\mathcal U_i,
 \label{eq:multi-stage-fusion}
\end{equation}
where $w_{iv}$ combines a fixed per-view reliability with alignment and boundary
quality, and the denominator renormalizes when preparation dropped a view.
Fusion precedes the analytical and graph constraints
so that their contributions remain separately measurable
(Table~\ref{tab:stage-results}).

\subsection{Architecture-Constrained Recovery}
\label{sec:constraint-filtering}

\parh{Analytical constraints.}
The fused list may still contain embedding collisions. ZKCache checks each
convolution key against Equations~\ref{eq:conv-output},
\ref{eq:conv-observables}, \ref{eq:window-overlap}, and \ref{eq:advice-volume}.
Nonintegral output dimensions, invalid padding, and incompatible groups are
rejected outright.  A key that survives is then scored against the trace.
Writing $\hat\psi^{(c)}(i)$ for the auxiliary estimates of unit-signature
duration, repetition profile, window overlap, and assignment-block volume taken
from position $i$, their disagreement with the key's source descriptors gives the
analytical-constraint penalty
\begin{equation}
 C_{eq}(q\mid i)=\sum_{c\in\mathcal B}
 \min\Bigl(\bigl\|\Lambda_c\bigl(\hat\psi^{(c)}(i)-\psi^{(c)}(h)\bigr)
 \bigr\|_1\big/\kappa_c,\;1\Bigr).
 \label{eq:analytical-penalty}
\end{equation}
For non-convolution keys, $C_{eq}$ applies the generic
operator-validity and shape checks, such as ReLU preserving tensor size and
pooling obeying Equation~\ref{eq:conv-output}, and is zero when no
operator-specific relation is available. These estimates are coarse, so they
contribute a penalty and do not replace learned retrieval. No relation is applied
to a single probe round.

\parh{Graph constraints.}
Graph constraints propagate tensor shapes between nodes.  Producer and consumer
shapes must agree after an explicitly allowed reshape or flatten; ReLU and Add
preserve shape, where Add is the elementwise addition that merges a residual
shortcut into the main path; pooling changes the spatial grid; and both inputs to
a residual Add must match.  Invalid edges, cycles, and unmatched merges are
rejected. If a position's key is absent from the database
(Table~\ref{tab:target-statistics}), rejecting all shape disagreements could
leave no feasible graph. $C_{graph}(G,\mathbf q)$ therefore counts admitted
shape disagreements at a cost, and $\Omega(G)$ counts nodes and edges that the
decoder adds beyond those supported by retained candidates. For a graph $G$ with
node set $V$ and node keys
$\mathbf q=(q_1,\ldots,q_{|V|})$, the decode cost is
\begin{equation}
 \begin{aligned}
 \Phi(G,\mathbf q)=
   &\sum_i D_{fuse}(q_i\mid i)+\lambda_{eq}\sum_i C_{eq}(q_i\mid i)\\
   &+\lambda_g C_{graph}(G,\mathbf q)+\mu\,\Omega(G).
 \end{aligned}
 \label{eq:decode-cost}
\end{equation}

\parh{Decoding.}
Decoding searches over identity and projection shortcuts by beam search,
rejecting cycles and incompatible Add inputs~\cite{he2016deep}.  When the
recovered operator sequence contains no merge operator, the graph can only be a
linear chain, and ZKCache falls back to a dynamic program that propagates tensor
shapes along it.  Add is a profiled operator type retrieved like any other
(Section~\ref{sec:retrieval-fusion}), so the choice follows from the recovered
types rather than from prior knowledge of the victim.

\parh{Output.}
ZKCache sorts the feasible hypotheses $R_j=(G_j,\mathbf q_j)$ by
Equation~\ref{eq:decode-cost} and emits $R_1$, the feasible architecture of
minimum cost.  It never abstains, so every mistake counts as a top-1 error.
Whether the source-derived dependence of Section~\ref{sec:leakage-analysis}
survives cache-set aggregation is an empirical question. Section~\ref{sec:evaluation}
evaluates it for four system--model pairs and measures the contribution of each
mechanism.

	\section{Evaluation}
\label{sec:evaluation}

This section evaluates ZKCache end to end.  We first describe the experimental
setup in Section~\ref{sec:experimental-setup} and define the recovery metrics in
Section~\ref{sec:evaluation-metrics}.  Section~\ref{sec:architecture-recovery}
then reports operator-type, hyperparameter, and exact-topology recovery, and
Section~\ref{sec:ablation} measures the contribution of each view and
constraint.

\subsection{Experimental Setup}
\label{sec:experimental-setup}

We test VGG16 and ResNet-18 on both EZKL and TeleSparse, giving the four
system--model pairs in Table~\ref{tab:target-statistics}.  For each pair, we fix
the input shape, quantization setting, compiled circuit, and proving
configuration.  Table~\ref{tab:platform-config} in
Appendix~\ref{app:platform-config} reports the complete hardware, software, and
core-placement configuration.  We run 30
independent trials per pair.
Each trial uses a fresh input and a process restart, and the attack receives a
single trace.  Each recovery-accuracy entry is the mean of the 30 per-trial
percentages.  Component comparisons assign equal weight to the four pairs.

We construct the shadow corpus as described in
Section~\ref{sec:shadow-dataset}.  VGG16 and ResNet-18 are excluded at the graph
level, although their operator configurations may occur in other source
models.  Table~\ref{tab:target-statistics} therefore states the resulting
database coverage.  Profiling, calibration, and shadow validation use disjoint
splits.  Before examining any target trace, we freeze the monitored cache sets,
normalization, encoder and loss hyperparameters, $k$, fusion and constraint
weights, and beam width.  EZKL and TeleSparse have separate traces, two-view model stacks,
embedding databases, and cache-set rankings.  They share only the source
descriptors, the analytical constraints, and the graph solvers.

\FloatBarrier
\begin{table}[!ht]
\vspace{-3pt}
\centering
\scriptsize
\setlength{\tabcolsep}{2.2pt}
\caption{Target workloads and database coverage.  Coverage is the
instance-weighted percentage of target operator positions whose complete
configuration key appears in the frozen embedding database.}
\label{tab:target-statistics}
\begin{tabular}{@{}llrrrr@{}}
\toprule
System & Model & Ops. & HP fields & Edges & Coverage \\
\midrule
EZKL & VGG16 & 37 & 113 & 36 & 97.3\% \\
EZKL & ResNet-18 & 49 & 147 & 56 & 95.9\% \\
TeleSparse & VGG16 & 37 & 113 & 36 & 94.6\% \\
TeleSparse & ResNet-18 & 49 & 147 & 56 & 91.8\% \\
\bottomrule
\end{tabular}
\vspace{-4pt}
\end{table}

\subsection{Recovery Metrics}
\label{sec:evaluation-metrics}

Let the ground-truth and predicted sequences contain $N$ and $\hat N$
operators.  We construct a one-to-one alignment $\mathcal M$ between detected
segments and instrumented intervals using a predefined, monotone
temporal-overlap rule.  Operator labels do not affect the alignment.
Operator-type accuracy is
\begin{equation}
 A_{op}=100\frac{\sum_{(i,j)\in\mathcal M}
 \mathbf{1}[\hat t_j=t_i]}{\max(N,\hat N)}.
 \label{eq:operator-accuracy}
\end{equation}
Every missing or extra operator therefore lowers the score.  Let
$\mathcal H_i$ be the preregistered set of fields for operator $i$, such as
kernel size, channels, stride, padding, or groups.  Hyperparameter
accuracy is
\begin{equation}
 A_{hp}=100\frac{\sum_{(i,j)\in\mathcal M}\mathbf{1}[\hat t_j=t_i]
 \sum_{f\in\mathcal H_i}\mathbf{1}[\hat h_{j,f}=h_{i,f}]}
 {\sum_i|\mathcal H_i|}.
 \label{eq:hyperparameter-accuracy}
\end{equation}
A missing or mistyped operator makes all of its applicable fields incorrect;
parameter-free operators contribute only to $A_{op}$.  The exact-topology
recovery rate (ETR) is the percentage of trials in which the complete
operator-type sequence and all graph edges match, irrespective of whether every
hyperparameter tuple is covered by the prototype database.  All metrics use the
final top-1 prediction.  Database misses remain in the denominator and are scored against
the architecture returned by ZKCache.

A tuple-valued quantity, such as a two-dimensional kernel, stride, or padding,
counts as one field and is correct only if all components match.  The 37-node VGG16 graph
contains 13 Conv, 15 ReLU, five MaxPool, one flatten, and three FC nodes.  The
49-node ResNet-18 graph contains 20 Conv, 17 ReLU, eight Add, one MaxPool, one
global-average-pool, one flatten, and one FC node.  Its projection branches
bring the total to 56 directed edges.

\subsection{Architecture Recovery Accuracy}
\label{sec:architecture-recovery}

Table~\ref{tab:graph-results} reports end-to-end recovery for each
system--model pair.  Operator-type and hyperparameter accuracy are the primary
results.  ETR additionally requires the complete recovered operator-type
sequence and all graph edges to match.

\FloatBarrier
\begin{table}[!ht]
\vspace{-3pt}
\centering
\scriptsize
\setlength{\tabcolsep}{2.4pt}
\caption{End-to-end top-1 architecture recovery. The first two rows report means of per-trial accuracies; ETR is the fraction of trials with an exact topology match.}
\label{tab:graph-results}
\begin{tabular}{@{}lcccc@{}}
\toprule
Metric & \shortstack{EZKL\\VGG16} & \shortstack{EZKL\\ResNet-18} &
\shortstack{TeleSparse\\VGG16} & \shortstack{TeleSparse\\ResNet-18} \\
\midrule
Operator-type accuracy (\%) & 99.4 & 98.8 & 98.5 & 97.7 \\
Hyperparameter accuracy (\%) & 95.6 & 94.1 & 92.7 & 90.4 \\
Exact topology rate (\%, aux.) & 93.3 & 86.7 & 90.0 & 80.0 \\
\bottomrule
\end{tabular}
\end{table}

Across the four pairs, the equal-weight averages are 98.6\% for operator types
and 93.2\% for hyperparameters.  ZKCache recovers the exact topology in 26 of 30
EZKL ResNet-18 trials and 24 of 30 TeleSparse ResNet-18 trials, for 50 of 60 trials
(83.3\%) overall.  TeleSparse has both lower profiled-key coverage and greater
sparsity-dependent trace variation, which coincide with its lower recovery
rates.  Uncovered keys remain in the hyperparameter-accuracy denominator but do
not by themselves prevent an ETR success.

These results use automatically detected boundaries and the full attack
pipeline.  Decoding follows the rule of
Section~\ref{sec:constraint-filtering} in every trial.  The recovered ResNet-18
sequences contain merge operators, so they are decoded by residual beam search,
whereas the recovered VGG16 sequences contain none and reduce to the sequential
dynamic program.
Each system uses its own two-view model stack and embedding database; we do not
apply EZKL embeddings to TeleSparse or claim portability across binary
representations.

\subsection{Impact of Views and Constraints}
\label{sec:ablation}

Table~\ref{tab:stage-results} measures how each source of evidence changes the
final operator-type and hyperparameter accuracy.  Duration-only and
handcrafted matching provide feature baselines. The next rows test the second
view, pooling-head specialization, encoder sharing, and source-derived
supervision before constraints are applied.  The single-head variant maps the
shared synthesis-composite feature sequence to one embedding.  In the untied
control, both encoders process the complete synthesis-composite segment but do
not share weights; neither variant separates tensor access and assignment into
independent stages.  All learned variants use the same profiled-key
database and a candidate budget of $k=5$.  The oracle-segmentation row changes
only the view boundaries and gives an upper bound rather than a
primary attack result.

The shared dual-head representation improves on a single pooling head by 3.8
and 11.4 points for operator types and hyperparameters, while sharing the
synthesis-composite encoder adds 1.6 and 5.5 points over untied encoders.
Source supervision and the analytical and graph constraints then raise the
full representation from 86.7/85.8 to 98.6/93.2, showing that learned local
evidence and explicit consistency are complementary.

\FloatBarrier
\begin{table}[!t]
\centering
\scriptsize
\setlength{\tabcolsep}{2.4pt}
\caption{Component ablation for accuracy.  Values are equal-weight
averages over the four system--model pairs.}
\label{tab:stage-results}
\resizebox{\columnwidth}{!}{%
\begin{tabular}{@{}lcc@{}}
\toprule
Method & Operator acc. (\%) & Hyperparam. acc. (\%) \\
\midrule
Duration only & 58.2 & 36.4 \\
Handcrafted matcher & 65.5 & 44.9 \\
Generic contrastive, dummy-layout only & 72.3 & 53.8 \\
Generic contrastive, synthesis-composite only & 77.8 & 61.7 \\
Two views, single synthesis-composite pooling head & 81.6 & 68.9 \\
Two views, dual pooling heads, untied encoders & 83.8 & 74.8 \\
Two views, shared encoder + dual pooling heads & 85.4 & 80.3 \\
Prover-aware full representation & 86.7 & 85.8 \\
\quad + analytical constraints & 97.4 & 90.9 \\
\quad + graph constraints (\textbf{ZKCache}) & \textbf{98.6} & \textbf{93.2} \\
\midrule
ZKCache with oracle segmentation & 99.2 & 94.5 \\
\bottomrule
\end{tabular}
}
\vspace{-4pt}
\end{table}
\FloatBarrier

Oracle boundaries add only 0.6 and 1.3 points for operator types and
hyperparameters; the remaining 0.8- and 5.5-point gaps mainly reflect similar
or absent database configurations.

\FloatBarrier
\raggedbottom

	\section{Defense: Budgeted Block Envelopes}
\label{sec:defense}

The evaluation shows that a single trace allows ZKCache to recover 93.2\% of
hyperparameter fields on average across the four system--model pairs.  For
convolution layers, exact tuples are exposed by parameter-dependent traversal
in both temporal views and by interleaved assignment in the
\emph{synthesis-composite} view.  This motivates a defense that removes
tuple-specific trace patterns under a bounded performance cost.

\begin{figure*}[t]
\centering
\includegraphics[width=0.95\textwidth]{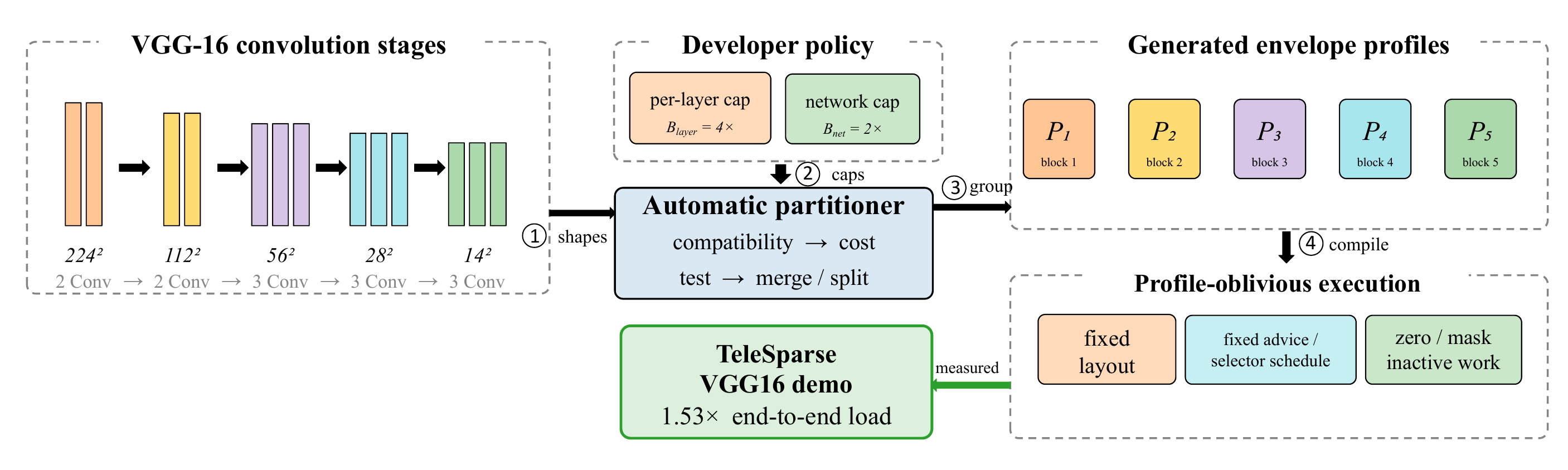}
\caption{Developer-budgeted envelope generation.  A compact VGG16 schematic
supplies layer shapes, developer caps drive automatic grouping, and every
resulting profile compiles to a fixed observable schedule.}
\label{fig:defense-auto-grouping}
\end{figure*}

Padding every convolution to the network-wide maxima would hide these
differences, but would make each small layer pay the cost of the largest one.
Instead, the developer sets an overhead budget, and the compiler automatically
groups compatible convolutions.  Each group receives a block-level envelope
with fixed padded bounds and a fixed schedule, so an observed trace reveals the
envelope rather than the layer's exact tuple.  Figure~\ref{fig:defense-auto-grouping}
summarizes this developer-controlled grouping and profile-oblivious execution
workflow.

\parh{Defense scope.}
Block envelopes directly protect convolution hyperparameters.  Their effect
also extends beyond convolution: channel and spatial dimensions constrain
adjacent pooling and FC keys as well as branch/merge compatibility during graph
decoding (Section~\ref{sec:constraint-filtering}).  Hiding these dimensions
removes graph-wide anchors and therefore weakens the recovery of
non-convolution parameters.

We do not attempt to hide operator types.  Different operators execute distinct
gadget and traversal paths; making them indistinguishable would require a
superset schedule that executes and masks every supported path.  The cost of
such a universal-circuit-like construction falls outside our convolution
work-proxy and developer-budget model.  Hiding them is
orthogonal future work.

\parh{Developer-controlled objective.}
For a convolution layer $\ell$, we use the source-derived work proxy
\begin{equation}
  W(\ell)=C_{out}H_oW_o\frac{C_{in}}{g}K_hK_w.
  \label{eq:defense-real-cost}
\end{equation}
For a candidate group $G$, profile $P(G)$ takes the component-wise maxima of
its members' input/output channels, output dimensions, kernel extent, and
padding.  Stride, group mode, bias handling, and successor semantics
must remain compatible.  Let $\widehat W(\ell;P(G))$ be the work incurred when
$\ell$ executes the full profile.  For a partition $\mathcal P$, we define
\begin{equation}
\begin{split}
 \Omega_{\max}(\mathcal P)
   &=\max_{G\in\mathcal P,\ell\in G}
     \frac{\widehat W(\ell;P(G))}{W(\ell)},\\
 \Omega_{\mathrm{net}}(\mathcal P)
   &=\frac{\sum_{G\in\mathcal P}\sum_{\ell\in G}
       \widehat W(\ell;P(G))}
      {\sum_{\ell}W(\ell)}.
 \label{eq:defense-overhead}
\end{split}
\end{equation}
The developer supplies a per-layer cap $B_{\mathrm{layer}}$ and a network-wide
cap $B_{\mathrm{net}}$.  The compiler then minimizes the number of observable
profiles subject to
\begin{equation}
 \min_{\mathcal P}|\mathcal P|\quad\text{s.t.}\quad
 \Omega_{\max}\le B_{\mathrm{layer}},\qquad
 \Omega_{\mathrm{net}}\le B_{\mathrm{net}}.
 \label{eq:defense-partition}
\end{equation}
Larger caps permit fewer, broader profiles and stronger parameter hiding;
tighter caps preserve performance by producing more, narrower profiles.

\parh{Automatic grouping.}
The pass first lists convolution layers in topological order and forms hard
compatibility buckets using output resolution, stride,
operator type, and graph-successor semantics.  Within each bucket, it attempts
shape-similar merges and recomputes both quantities in
Equation~\ref{eq:defense-overhead}.  A merge is accepted only while both
developer caps remain satisfied; otherwise the candidate is split. The output is a small profile table and a static schedule for
each profile.

\parh{Oblivious profile execution.}
Every member executes the full channel, kernel, and output ranges of its
profile.  Inactive channels, kernel taps, and output positions are represented
by zeros and masks, but no data-dependent branch may skip them.  Consequently,
dummy layout, tensor traversal, and advice/selector assignment retain a fixed
profile-specific shape.  The defense intentionally leaks the much coarser
profile identifier while suppressing the exact convolution tuple within~it.

\parh{Prototype result.}
Our TeleSparse VGG16 prototype uses developer settings
$B_{\mathrm{layer}}=4\times$ and $B_{\mathrm{net}}=2\times$.  The compiler
automatically derives the envelope groups, and the protected prover reaches a
normalized end-to-end load of $1.53\times$ relative to the unprotected run,
i.e., a 53\% increase while remaining below the requested network budget.  The
number is a measured design point rather than a universal constant: developers
can tighten the caps for lower overhead or relax them to merge more layers and
hide parameters within broader profiles.

\parh{Security effect.}
To quantify the defense, we rerun ZKCache against the protected
TeleSparse VGG16 prover under the same single-trace recovery and
hyperparameter metric used in Section~\ref{sec:evaluation}.  Hyperparameter
recovery accuracy falls from $92.7\%$ on the unprotected prover to $24.6\%$ on
the protected prover, a reduction of 68.1 percentage points.  Because this
metric covers every preregistered field of each supported parameterized
operator, the drop reflects both direct hiding of convolution tuples and the
loss of precise convolution anchors used to disambiguate other operators
through graph constraints.

	\section{Discussion and Limitations}
\label{sec:discussion}

\parh{Protocol privacy versus execution privacy.}
ZKCache does not break Halo2 soundness or extract information from a proof
transcript.  It demonstrates that a relation whose implementation depends on
a secret description can leak that description while constructing a
zero-knowledge proof.  Constant-time finite-field code is therefore
necessary but insufficient for architecture-private ZKML.  Conversely, our
findings do not weaken deployments where circuit shape is intentionally public.


\parh{Weights and layouts.}
ZKCache recovers tensor dimensions and graph structure, not numerical weights.
Physical weight layout matters to which cache sets are active and to attack
calibration, but the current EZKL convolution canonicalizes formats and slices
one output-channel kernel at a time.  Our parameter equations depend on the
logical access count, not on knowing exact weight addresses.  Recovering
weights, sparsity masks, or quantized values would require a different
data-dependent leakage analysis, such as memory-side extraction techniques
designed specifically for model weights~\cite{rakin2022deepsteal}.

\parh{Portability.}
ZKCache's prover-aware methodology can be adapted to other PLONKish ZKML
implementations when circuit sizing or synthesis retains model-dependent
execution, but the concrete temporal boundaries and learned trace stack remain
implementation-specific. Our TeleSparse results show that the workflow remains
effective after system-specific profiling; they do not demonstrate cross-binary
embedding transfer. STARK- and GPU-based provers are outside our evaluation and
would require new analysis of their execution schedules, collectors, and learned
models.

\parh{Limitations.}
Prime+Probe resolution is lower than shared-memory attacks and depends on
placement, stable scheduling, and profiling.  Prefetchers and allocators can
both create and erase signatures.  Contrastive training can reduce these
effects but cannot guarantee allocation invariance and may learn shortcuts such
as duration, process placement, or system-specific noise.  Exact
recovery also remains bounded by database coverage: an unseen parameter tuple is not
created by nearest-neighbor retrieval.
Graphs outside the supported residual grammar may be unrecoverable or leave
several feasible topologies.  ETR counts a trial as successful only when the
selected top-1 operator-type sequence and all graph edges are~exact.

	\section{Related Work}
\label{sec:related-work}

\parh{DNN architecture side channels.}
Prior work infers layer sequences from shared framework calls, recovers matrix
dimensions from tiled GEMM, or maps compiled-executable cache traces to
operators and
hyperparameters~\cite{hong2018security,cachetelepathy2020,deepcache2024}.  Representation-based
attacks further learn repeated trace structure and match it against profiled
operator databases~\cite{deepcache2024}.  ZKCache instead models two
ZKML-specific temporal views and combines learned local evidence with
source-derived and graph constraints.

\parh{Zero-knowledge machine learning.}
Prior ZKML systems established verifiable inference and improved conversions,
convolution proofs, and private-model
support~\cite{ghodsi2017safetynets,weng2021mystique,liu2021zkcnn}.  Modern systems
compile quantized inference into arithmetic constraints while balancing cost,
accuracy, and expressiveness~\cite{chen2024zkml,ezkl}; TeleSparse reduces Halo2
cost through sparsification and neural teleportation~\cite{telesparse2025}.
Whereas this line emphasizes protocol security and performance, we ask whether
CPU proof generation leaks a private architecture.

\parh{Side channels in zero-knowledge implementations.}
Implementation leakage can persist despite secure protocol definitions.
Circuit-level work identifies under-constrained or inconsistent ZK
implementations~\cite{pailoor2023automated,wen2023practical}, while recent cache
analysis targets secrets in finite-field libraries and ZK-friendly hash
implementations~\cite{mukherjee2025cachezk}.  ZKCache targets a different layer:
model shapes govern tensor traversal, circuit rows, and assignment schedules
even when field operations are constant-time.

	\section{Conclusion}
\label{sec:conclusion}

We presented ZKCache, a two-view cache attack on Halo2-based ZKML provers.  It
models dummy-layout and synthesis-composite execution separately, uses one
shared synthesis encoder to extract interleaved tensor-access and assignment
evidence, and applies source-derived and graph-wide constraints to recover the
architecture.  Across EZKL and TeleSparse, ZKCache achieves 98.6\% operator-type
and 93.2\% hyperparameter accuracy, and recovers ResNet-18's complete
operator-type sequence and residual wiring in 83.3\% of trials.  Our budgeted
block envelopes reduce hyperparameter
recovery from 92.7\% to 24.6\% at $1.53\times$ total end-to-end load.  These
results show that transcript zero knowledge does not imply an execution-oblivious
prover; private-architecture deployments must treat both temporal views as part
of their side-channel boundary.

	\section{Ethics Considerations}
\label{sec:ethics}

ZKCache is dual-use: it helps developers identify architecture leakage, but it
could also assist a co-resident adversary in recovering proprietary model
structure.  We evaluated the attack only on locally deployed EZKL and
TeleSparse instances using public model architectures; we did not probe
third-party production services or collect personal data.  This work involved
no human participants or human-derived data.

\parh{Responsible disclosure.}
We disclosed the vulnerability details to the EZKL and TeleSparse development
teams before publication.  Both teams acknowledged our findings, and we are
actively coordinating with them on mitigation strategies.  We also present the
budgeted block-envelope defense in Section~\ref{sec:defense}, allowing developers
to reduce exact hyperparameter leakage under a deployment-specific performance
budget.

	\bibliographystyle{IEEEtran}
	\bibliography{bib/machine-learning,bib/timing,bib/sidechannel,bib/analysis,bib/ref,bib/testing-cv,bib/cv,bib/sw,bib/zkcache-additions}

@misc{simonyan2014very,
  title={Very deep convolutional networks for large-scale image recognition},
  author={Simonyan, Karen and Zisserman, Andrew},
  howpublished={arXiv preprint arXiv:1409.1556},
  url={https://arxiv.org/abs/1409.1556},
  year={2014}
}

@inproceedings{he2016deep,
  title={Deep residual learning for image recognition},
  author={He, Kaiming and Zhang, Xiangyu and Ren, Shaoqing and Sun, Jian},
  booktitle={Proceedings of the IEEE conference on computer vision and pattern recognition},
  pages={770--778},
  publisher={IEEE Computer Society},
  address={Los Alamitos, CA, USA},
  year={2016}
}

@inproceedings{10.1145/22145.22178,
author = {Goldwasser, S and Micali, S and Rackoff, C},
title = {The Knowledge Complexity of Interactive Proof-Systems},
year = {1985},
isbn = {0897911512},
publisher = {Association for Computing Machinery},
address = {New York, NY, USA},
url = {https://doi.org/10.1145/22145.22178},
doi = {10.1145/22145.22178},
booktitle = {Proceedings of the Seventeenth Annual ACM Symposium on Theory of Computing},
pages = {291--304},
numpages = {14},
location = {Providence, Rhode Island, USA},
series = {STOC '85}
}

@article{pailoor2023automated,
  title={Automated Detection of Under-Constrained Circuits in Zero-Knowledge Proofs},
  author={Pailoor, Shankara and Chen, Yanju and Wang, Franklyn and Rodr{\'\i}guez, Clara and Van Geffen, Jacob and Morton, Jason and Chu, Michael and Gu, Brian and Feng, Yu and Dillig, Isil},
  journal={Proceedings of the ACM on Programming Languages},
  volume={7},
  number={PLDI},
  pages={1510--1532},
  year={2023},
  publisher={ACM},
  doi={10.1145/3591282}
}

@inproceedings{wen2023practical,
  title={Practical Security Analysis of Zero-Knowledge Proof Circuits},
  author={Wen, Hongbo and Stephens, Jon and Chen, Yanju and Ferles, Kostas and Pailoor, Shankara and Charbonnet, Kyle and Dillig, Isil and Feng, Yu},
  booktitle={33rd USENIX Security Symposium (USENIX Security 24)},
  pages={1471--1487},
  year={2024},
  publisher={USENIX Association},
  address={Philadelphia, PA, USA},
  url={https://www.usenix.org/conference/usenixsecurity24/presentation/wen}
}

@misc{cryptoeprint:2019/953,
      author = {Ariel Gabizon and Zachary J.  Williamson and Oana Ciobotaru},
      title = {PLONK: Permutations over Lagrange-bases for Oecumenical Noninteractive arguments of Knowledge},
      howpublished = {Cryptology ePrint Archive, Paper 2019/953},
      year = {2019},
      url = {https://eprint.iacr.org/2019/953}
}

@inproceedings{liu2015last,
  title={Last-level cache side-channel attacks are practical},
  author={Liu, Fangfei and Yarom, Yuval and Ge, Qian and Heiser, Gernot and Lee, Ruby B},
  booktitle={2015 IEEE symposium on security and privacy},
  pages={605--622},
  year={2015},
  organization={IEEE},
  publisher={IEEE Computer Society},
  address={Los Alamitos, CA, USA}
}

@inproceedings{osvik2006cache,
 author = {Osvik, Dag Arne and Shamir, Adi and Tromer, Eran},
 title = {Cache Attacks and Countermeasures: The Case of {AES}},
 booktitle = {Topics in Cryptology -- CT-RSA 2006},
 series = {Lecture Notes in Computer Science},
 volume = {3860},
 pages = {1--20},
 publisher = {Springer},
 address = {Berlin, Heidelberg},
 doi = {10.1007/11605805_1},
 year = {2006},
}

@misc{hong2018security,
  title={Security analysis of deep neural networks operating in the presence of cache side-channel attacks},
  author={Hong, Sanghyun and Davinroy, Michael and Kaya, Yi{\u{g}}itcan and Locke, Stuart Nevans and Rackow, Ian and Kulda, Kevin and Dachman-Soled, Dana and Dumitra{\c{s}}, Tudor},
  howpublished={arXiv preprint arXiv:1810.03487},
  url={https://arxiv.org/abs/1810.03487},
  year={2018}
}

@inproceedings{oren2015spy,
  title={The Spy in the Sandbox: Practical Cache Attacks in {JavaScript} and Their Implications},
  author={Oren, Yossef and Kemerlis, Vasileios P and Sethumadhavan, Simha and Keromytis, Angelos D},
  booktitle={Proceedings of the 22nd ACM SIGSAC Conference on Computer and Communications Security},
  pages={1406--1418},
  year={2015},
  publisher={ACM},
  address={New York, NY, USA},
  doi={10.1145/2810103.2813708}
}

@inproceedings{hong20200wn,
title={How to 0wn {NAS} in Your Spare Time},
author={Sanghyun Hong and Michael Davinroy and Yi{\u{g}}itcan Kaya and Dana Dachman-Soled and Tudor Dumitra{\c{s}}},
booktitle={International Conference on Learning Representations},
year={2020},
publisher={OpenReview.net},
address={Amherst, MA, USA},
numpages={17},
url={https://openreview.net/forum?id=S1erpeBFPB}
}

@inproceedings{ganred2020,
  author    = {Liu, Yuntao and Srivastava, Ankur},
  title     = {{GANRED}: {GAN}-Based Reverse Engineering of {DNNs} via Cache
               Side-Channel},
  booktitle = {Proceedings of the 2020 ACM SIGSAC Conference on Cloud Computing
               Security Workshop},
  pages     = {41--52},
  year      = {2020},
  publisher = {Association for Computing Machinery},
  address   = {New York, NY, USA},
  doi       = {10.1145/3411495.3421356}
}

@inproceedings{deepsniffer2020,
  author    = {Hu, Xing and Liang, Ling and Li, Shuangchen and Deng, Lei and
               Zuo, Pengfei and Ji, Yu and Xie, Xinfeng and Ding, Yufei and
               Liu, Chang and Sherwood, Timothy and Xie, Yuan},
  title     = {{DeepSniffer}: A {DNN} Model Extraction Framework Based on
               Learning Architectural Hints},
  booktitle = {Proceedings of the Twenty-Fifth International Conference on
               Architectural Support for Programming Languages and Operating
               Systems},
  pages     = {385--399},
  year      = {2020},
  publisher = {Association for Computing Machinery},
  address   = {New York, NY, USA},
  doi       = {10.1145/3373376.3378460}
}

@inproceedings{deepcache2024,
  author    = {Liu, Zhibo and Yuan, Yuanyuan and Chen, Yanzuo and Hu, Sihang
               and Li, Tianxiang and Wang, Shuai},
  title     = {{DeepCache}: Revisiting Cache Side-Channel Attacks in Deep
               Neural Networks Executables},
  booktitle = {Proceedings of the 2024 ACM SIGSAC Conference on Computer and
               Communications Security},
  pages     = {3632--3648},
  year      = {2024},
  publisher = {Association for Computing Machinery},
  address   = {New York, NY, USA},
  doi       = {10.1145/3658644.3690241}
}

@inproceedings{chen2024zkml,
  author    = {Chen, Bing-Jyue and Waiwitlikhit, Suppakit and Stoica, Ion and
               Kang, Daniel},
  title     = {{ZKML}: An Optimizing System for ML Inference in Zero-Knowledge
               Proofs},
  booktitle = {Proceedings of the Nineteenth European Conference on Computer
               Systems},
  pages     = {560--574},
  year      = {2024},
  publisher = {Association for Computing Machinery},
  address   = {New York, NY, USA},
  doi       = {10.1145/3627703.3650088}
}

@misc{ezkl,
  author       = {{Zkonduit}},
  title        = {{EZKL}: Easy Zero-Knowledge Inference},
  howpublished = {\url{https://github.com/zkonduit/ezkl}},
  year         = {2026}
}

@article{telesparse2025,
  author  = {Maheri, Mohammad M. and Haddadi, Hamed and Davidson, Alex},
  title   = {{TeleSparse}: Practical Privacy-Preserving Verification of Deep
             Neural Networks},
  journal = {Proceedings on Privacy Enhancing Technologies},
  volume  = {2025},
  number  = {4},
  pages   = {861--880},
  year    = {2025},
  doi     = {10.56553/popets-2025-0161}
}

@misc{halo2book,
  author       = {{Electric Coin Company}},
  title        = {The {Halo2} Book},
  howpublished = {\url{https://zcash.github.io/halo2/}},
  year         = {2026},
  note         = {Accessed August 5, 2026}
}

@misc{cryptoeprint:2020/315,
  author       = {Gabizon, Ariel and Williamson, Zachary J.},
  title        = {{Plookup}: A Simplified Polynomial Protocol for Lookup
                  Tables},
  howpublished = {Cryptology ePrint Archive, Paper 2020/315},
  year         = {2020},
  url          = {https://eprint.iacr.org/2020/315}
}

@inproceedings{cachetelepathy2020,
  author    = {Yan, Mengjia and Fletcher, Christopher W. and Torrellas, Josep},
  title     = {Cache Telepathy: Leveraging Shared Resource Attacks to Learn
               {DNN} Architectures},
  booktitle = {29th USENIX Security Symposium},
  pages     = {2003--2020},
  year      = {2020},
  publisher = {USENIX Association},
  address   = {Berkeley, CA, USA},
  url       = {https://www.usenix.org/conference/usenixsecurity20/presentation/yan}
}

@inproceedings{mukherjee2025cachezk,
  author = {Mukherjee, Shibam and Rechberger, Christian and Schofnegger, Markus},
  title = {Cache Timing Leakages in Zero-Knowledge Protocols},
  booktitle = {7th Conference on Advances in Financial Technologies (AFT)},
  series = {Leibniz International Proceedings in Informatics},
  volume = {354},
  pages = {1:1--1:26},
  publisher = {Schloss Dagstuhl -- Leibniz-Zentrum fuer Informatik},
  address = {Wadern, Germany},
  year = {2025},
  doi = {10.4230/LIPIcs.AFT.2025.1},
  url = {https://drops.dagstuhl.de/entities/document/10.4230/LIPIcs.AFT.2025.1}
}

@misc{onnxmodelzoo,
  author       = {{ONNX Community}},
  title        = {{ONNX Model Zoo}},
  howpublished = {\url{https://github.com/onnx/models}},
  year         = {2026},
  note         = {Accessed August 6, 2026},
}

@string{PLDI ={ACM SIGPLAN Conf. on Programming Language Design and Implementation (PLDI)}}

@string{st = "\textsuperscript{st} "}

@string{nd = "\textsuperscript{nd} "}

@string{rd = "\textsuperscript{rd} "}

@string{th = "\textsuperscript{th} "}

@inproceedings{Yarom14,
 author = {Yarom, Yuval and Falkner, Katrina},
 title = {{FLUSH+RELOAD}: A High Resolution, Low Noise, {L3} Cache Side-channel Attack},
 booktitle = {Proceedings of the 23rd USENIX Conference on Security Symposium},
 year = {2014},
 pages = {719--732},
 publisher = {USENIX Association},
 address = {Berkeley, CA, USA},
}

@inproceedings{batina2019csinn,
  author={Lejla Batina and Shivam Bhasin and Dirmanto Jap and Stjepan Picek},
  title={{CSI} {NN}: Reverse Engineering of Neural Network Architectures Through Electromagnetic Side Channel},
  booktitle={28th USENIX Security Symposium (USENIX Security 19)},
  pages={515--532}, year={2019}, publisher={USENIX Association}, address={Santa Clara, CA, USA},
  url={https://www.usenix.org/conference/usenixsecurity19/presentation/batina}
}

@inproceedings{yu2020deepem,
  author={Honggang Yu and Haocheng Ma and Kaichen Yang and Yiqiang Zhao and Yier Jin},
  title={{DeepEM}: Deep Neural Networks Model Recovery through {EM} Side-Channel Information Leakage},
  booktitle={2020 IEEE International Symposium on Hardware Oriented Security and Trust (HOST)},
  pages={209--218}, year={2020}, publisher={IEEE}, address={Piscataway, NJ, USA},
  doi={10.1109/HOST45689.2020.9300274}
}

@inproceedings{zhu2021hermes,
  author={Yuankun Zhu and Yueqiang Cheng and Husheng Zhou and Yantao Lu},
  title={Hermes Attack: Steal {DNN} Models with Lossless Inference Accuracy},
  booktitle={30th USENIX Security Symposium (USENIX Security 21)},
  pages={1973--1988}, year={2021}, publisher={USENIX Association}, address={Berkeley, CA, USA},
  url={https://www.usenix.org/conference/usenixsecurity21/presentation/zhu}
}

@inproceedings{wu2022dnd,
  author={Ruoyu Wu and Taegyu Kim and Dave (Jing) Tian and Antonio Bianchi and Dongyan Xu},
  title={{DnD}: A {Cross-Architecture} Deep Neural Network Decompiler},
  booktitle={31st USENIX Security Symposium (USENIX Security 22)},
  pages={2135--2152}, year={2022}, publisher={USENIX Association}, address={Boston, MA, USA},
  url={https://www.usenix.org/conference/usenixsecurity22/presentation/wu-ruoyu}
}

@inproceedings{liu2023decompiling,
  author={Zhibo Liu and Yuanyuan Yuan and Shuai Wang and Xiaofei Xie and Lei Ma},
  title={Decompiling x86 Deep Neural Network Executables},
  booktitle={32nd USENIX Security Symposium (USENIX Security 23)},
  pages={7357--7374}, year={2023}, publisher={USENIX Association}, address={Anaheim, CA, USA},
  url={https://www.usenix.org/conference/usenixsecurity23/presentation/liu-zhibo}
}

@misc{feng2021zen,
  author={Feng, Boyuan and Qin, Lianke and Zhang, Zhenfei and Ding, Yufei and Chu, Shumo},
  title={{ZEN}: An Optimizing Compiler for Verifiable, Zero-Knowledge Neural Network Inferences},
  year={2021}, howpublished={Cryptology ePrint Archive, Paper 2021/087},
  url={https://eprint.iacr.org/2021/087}
}

@inproceedings{hao2024nonlinear,
  author={Meng Hao and Hongfei Chen and Kai Li and Shengshan Weng and Runchen Zhang and Minghui Yang and Ge Zhang},
  title={Scalable Zero-Knowledge Proofs for Non-linear Functions in Machine Learning},
  booktitle={33rd USENIX Security Symposium (USENIX Security 24)},
  pages={3819--3836}, year={2024}, publisher={USENIX Association}, address={Philadelphia, PA, USA},
  url={https://www.usenix.org/conference/usenixsecurity24/presentation/hao-meng-scalable}
}

@inproceedings{purnal2021primescope,
  author={Antoon Purnal and Furkan Turan and Ingrid Verbauwhede},
  title={{Prime+Scope}: Overcoming the Observer Effect for High-Precision Cache Contention Attacks},
  booktitle={Proceedings of the 2021 ACM SIGSAC Conference on Computer and Communications Security},
  pages={2906--2920}, year={2021}, publisher={ACM}, address={New York, NY, USA},
  doi={10.1145/3460120.3484816}
}

@inproceedings{chen2018tvm,
  author={Tianqi Chen and Thierry Moreau and Ziheng Jiang and Lianmin Zheng and Eddie Yan and Haichen Shen and Meghan Cowan and Leyuan Wang and Yuwei Hu and Luis Ceze and Carlos Guestrin and Arvind Krishnamurthy},
  title={{TVM}: An Automated End-to-End Optimizing Compiler for Deep Learning},
  booktitle={13th USENIX Symposium on Operating Systems Design and Implementation (OSDI 18)},
  pages={578--594}, year={2018}, publisher={USENIX Association}, address={Carlsbad, CA, USA}
}

@inproceedings{lavin2016fast,
  author={Andrew Lavin and Scott Gray}, title={Fast Algorithms for Convolutional Neural Networks},
  booktitle={2016 IEEE Conference on Computer Vision and Pattern Recognition (CVPR)},
  pages={4013--4021}, year={2016}, publisher={IEEE}, address={Piscataway, NJ, USA}, doi={10.1109/CVPR.2016.435}
}

@article{sakoe1978dtw,
  author={Hiroaki Sakoe and Seibi Chiba},
  title={Dynamic Programming Algorithm Optimization for Spoken Word Recognition},
  journal={IEEE Transactions on Acoustics, Speech, and Signal Processing},
  volume={26}, number={1}, pages={43--49}, year={1978}, doi={10.1109/TASSP.1978.1163055}
}

@article{killick2012changepoint,
  author={Rebecca Killick and Paul Fearnhead and Idris A. Eckley},
  title={Optimal Detection of Changepoints with a Linear Computational Cost},
  journal={Journal of the American Statistical Association}, volume={107}, number={500},
  pages={1590--1598}, year={2012}, doi={10.1080/01621459.2012.737745}
}

@inproceedings{khosla2020supervised,
  author={Prannay Khosla and Piotr Teterwak and Chen Wang and Aaron Sarna and Yonglong Tian and Phillip Isola and Aaron Maschinot and Ce Liu and Dilip Krishnan},
  title={Supervised Contrastive Learning}, booktitle={Advances in Neural Information Processing Systems},
  volume={33}, pages={18661--18673}, year={2020}, publisher={Curran Associates, Inc.},
  address={Red Hook, NY, USA}
}

@inproceedings{snell2017prototypical,
  author={Jake Snell and Kevin Swersky and Richard Zemel}, title={Prototypical Networks for Few-shot Learning},
  booktitle={Advances in Neural Information Processing Systems}, volume={30}, pages={4077--4087}, year={2017},
  publisher={Curran Associates, Inc.}, address={Red Hook, NY, USA}
}

@inproceedings{ghodsi2017safetynets,
  author={Zahra Ghodsi and Tianyu Gu and Siddharth Garg},
  title={{SafetyNets}: Verifiable Execution of Deep Neural Networks on an Untrusted Cloud},
  booktitle={Advances in Neural Information Processing Systems}, volume={30}, pages={4672--4681}, year={2017},
  publisher={Curran Associates, Inc.}, address={Red Hook, NY, USA}
}

@inproceedings{weng2021mystique,
  author={Chenkai Weng and Kang Yang and Xiang Xie and Jonathan Katz and Xiao Wang},
  title={Mystique: Efficient Conversions for Zero-Knowledge Proofs with Applications to Machine Learning},
  booktitle={30th USENIX Security Symposium (USENIX Security 21)},
  pages={501--518}, year={2021}, publisher={USENIX Association}, address={Berkeley, CA, USA},
  url={https://www.usenix.org/conference/usenixsecurity21/presentation/weng}
}

@inproceedings{liu2021zkcnn,
  author={Tianyi Liu and Xiang Xie and Yupeng Zhang},
  title={{zkCNN}: Zero Knowledge Proofs for Convolutional Neural Network Predictions and Accuracy},
  booktitle={Proceedings of the 2021 ACM SIGSAC Conference on Computer and Communications Security},
  pages={2968--2985}, year={2021}, publisher={ACM}, address={New York, NY, USA}, doi={10.1145/3460120.3485379}
}

@inproceedings{rakin2022deepsteal,
  author={Adnan Siraj Rakin and Md Hafizul Islam Chowdhuryy and Fan Yao and Deliang Fan},
  title={{DeepSteal}: Advanced Model Extractions Leveraging Efficient Weight Stealing in Memories},
  booktitle={2022 IEEE Symposium on Security and Privacy (SP)}, pages={1157--1174},
  year={2022}, publisher={IEEE}, address={Piscataway, NJ, USA}, doi={10.1109/SP46214.2022.9833743}
}
	\appendices

\section{Detailed Two-View Operator Signatures}
\label{app:operator-signatures}

Table~\ref{tab:operator-signatures} expands the source-derived comparison
summarized by Figure~\ref{fig:operator-traces}.  Each entry describes one
complete operator interval; tensor access and cell assignment remain
fine-grained and interleaved within the synthesis-composite view $s$.

\begin{table}[!ht]
\centering
\scriptsize
\setlength{\tabcolsep}{3pt}
\caption{Detailed source-derived two-view signatures for four representative
operators in the evaluated EZKL revision.  The joint cue states why combining
$d$ and $s$ distinguishes each operator.}
\label{tab:operator-signatures}
\begin{tabular}{@{}L{0.13\columnwidth}L{0.10\columnwidth}L{0.70\columnwidth}@{}}
\toprule
Operator & Evidence & Source-derived signature \\
\midrule
\multirow{3}{*}{Conv}
  & $d$ & Spatial-window sweeps nested within filter epochs. \\
  & $s$ & Interleaved image/filter access, dot-product gadgets, optional bias,
          and cell assignment. \\
  & $d+s$ & Nested window sweeps plus recurring filter and dot-product work. \\
\addlinespace[2pt]
\multirow{3}{*}{MaxPool}
  & $d$ & Repeated window traversal without a filter epoch. \\
  & $s$ & Interleaved window access, sort, shuffle/permutation, comparison, and
          equality work. \\
  & $d+s$ & Conv-like window repetition, but no filter stream or dot-product
            pattern. \\
\addlinespace[2pt]
\multirow{3}{*}{ReLU}
  & $d$ & A few flat, elementwise passes without a window or filter hierarchy. \\
  & $s$ & Interleaved activation access, sign decomposition, equality masking,
          and multiplication. \\
  & $d+s$ & A flat traversal paired with a short gate chain. \\
\addlinespace[2pt]
\multirow{3}{*}{FC}
  & $d$ & Repeated non-spatial input/weight reductions. \\
  & $s$ & Interleaved operand access and matrix-contraction assignment. \\
  & $d+s$ & Long contractions without spatial-window displacement, typically
            with fewer repetitions. \\
\bottomrule
\end{tabular}
\end{table}
\FloatBarrier

\section{EZKL Convolution Layout Excerpt}
\label{app:conv-source}

Listing~\ref{lst:conv-source} provides the source excerpt supporting the
iteration order analyzed in Section~\ref{sec:conv-hyperparameter-leakage}.

\noindent\hspace*{0.03\columnwidth}%
\begin{minipage}{0.94\columnwidth}
\begin{lstlisting}[
  basicstyle=\ttfamily\scriptsize,
  columns=fullflexible,
  breaklines=true,
  breakatwhitespace=true,
  frame=single,
  numbers=none,
  linewidth=\linewidth,
  xleftmargin=0pt,
  xrightmargin=0pt,
  framexleftmargin=2.5pt,
  framexrightmargin=2.5pt,
  captionpos=b,
  abovecaptionskip=7pt,
  aboveskip=0.6\baselineskip,
  belowskip=0.5\baselineskip,
  caption={Abridged from EZKL's \texttt{conv} layout in
  \protect\path{src/circuit/ops/layouts.rs}~\cite{ezkl}.},
  label={lst:conv-source}]
let mut iterations = vec![
    0..batch_size,
    0..num_groups,
    0..output_channels_per_group,
];
for slide in slides.iter() {
    iterations.push(0..*slide);
}

let coord = cartesian_coord_per_group[batch_offset + i]
    * stride;
slices.push(coord..(coord + kernel_dims[2 + i]));
let mut local_image = working_image.get_slice(&slices)?;
local_image.flatten();

let start = group * output_channels_per_group + i;
let mut local_kernel = working_kernel
    .get_slice(&[start..start + 1])?;
local_kernel.flatten();
let res = dot(config, region,
    &[&local_image, &local_kernel])?;
\end{lstlisting}
\end{minipage}

\FloatBarrier
\newpage
\section{Hardware and Attack Configuration}
\label{app:platform-config}

Table~\ref{tab:platform-config} fixes the execution environment used for both
shadow profiling and victim measurements.  The eviction-set construction is
timing-based, following the conflict-testing principle of
Prime+Scope~\cite{purnal2021primescope}; it does not read physical~addresses.

\begin{table}[!ht]
\centering
\scriptsize
\setlength{\tabcolsep}{3pt}
\renewcommand{\arraystretch}{1.06}
\caption{Hardware, software, deployment, and collection configuration.}
\label{tab:platform-config}
\begin{tabular}{@{}L{0.27\columnwidth}L{0.66\columnwidth}@{}}
\toprule
Item & Configuration \\
\midrule
Processor & Intel Core i7-9700K (Coffee Lake Refresh), 8 physical cores/8
threads, 3.60\,GHz base frequency. \\
Cache geometry & 32\,KiB 8-way L1D and 256\,KiB 4-way L2 per core; 12\,MiB
LLC with 8 slices, 2,048 sets per slice, 12 ways per slice, and 64-byte lines. \\
Memory & 32\,GiB dual-channel DDR4-2666. \\
Software & Ubuntu 22.04.5 LTS, Linux 5.15.0-130-generic, GCC 11.4.0,
Rust 1.83.0, and Python 3.10.12. \\
Deployment & Bare metal; no VM or container.  Shadow and target provers run on
the same host.  The attacker is an unprivileged user process. \\
Frequency controls & SMT is unavailable on this CPU.  Turbo Boost is disabled;
\texttt{intel\_pstate} uses the performance governor at 3.60\,GHz, and package
C-states are limited to C1. \\
Prefetchers & L1 and L2 hardware prefetchers remain enabled at their BIOS
defaults; the permuted probe order is retained in all runs. \\
Core placement & The victim is pinned to core 2 and the attacker to core 6
using \texttt{sched\_setaffinity}.  They occupy different physical cores on
the same package and share only the LLC; neither is an SMT sibling. \\
Timing & Fenced \texttt{RDTSCP} timestamps; 2,000-cycle Prime+Probe wait
(0.56\,$\mu$s at 3.60\,GHz); probe order is freshly permuted per sample. \\
Eviction sets & A 1\,GiB anonymous buffer requests 2\,MiB transparent huge
pages with \texttt{madvise(MADV\_HUGEPAGE)}.  Timing-only group reduction finds
12 congruent lines per slice and retains sets with at least 95\% measured
eviction probability.  No root privilege, \texttt{/proc/pagemap}, or physical
address information is used. \\

\bottomrule
\end{tabular}
\end{table}

\end{document}